\documentclass{aa}  
\usepackage{graphicx}
\usepackage{txfonts}
\begin{document}

   \title{Presence of dust with a UV bump in massive, star-forming galaxies 
          at $1 < z < 2.5$}

   \author{S. Noll\inst{1}
   \and D. Pierini\inst{1}
   \and M. Pannella\inst{1}
   \and S. Savaglio\inst{1}}

   \offprints{S. Noll}

   \institute{Max-Planck-Institut f\"ur extraterrestrische Physik, 
              Giessenbachstr., 85748 Garching, Germany\\
              \email{snoll,dpierini,maurilio,savaglio@mpe.mpg.de}
             }

   \date{Accepted 4 July 2007}

 
  \abstract
   {}
{Fundamental properties of the extinction curve, like the slope in the 
rest-frame ultraviolet (UV) and the presence/absence of a broad absorption 
excess centred at 2175\,\AA{} (the UV bump), are investigated for a sample of 
108 massive, star-forming galaxies at $1 < z < 2.5$, selected from the FDF 
Spectroscopic Survey, the K20 survey, and the GDDS.}
{These characteristics are constrained from a parametric description of the 
UV spectral energy distribution (SED) of a galaxy, as enforced by combined 
stellar population and radiative transfer models for different geometries, 
dust/stars configurations and dust properties.}
{In at least one third of the sample, there is a robust evidence for 
extinction curves with at least a moderate UV bump. The presence of the 
carriers of the UV bump is more evident in galaxies with UV SEDs suffering 
from heavy reddening. We interpret these results as follows. The sample 
objects possess different mixtures of dust grains and molecules producing 
extinction curves in between the average ones of the Small and Large 
Magellanic Cloud, where the UV bump is absent or modest, respectively. Most 
of the dust embeds the UV-emitting stellar populations or is distributed out 
of the galaxy mid-plane. Alternatively, even dust with a pronounced UV bump,
as for the average Milky-Way extinction curve, can be present and distributed 
in the galaxy mid-plane. In this case, variations of the continuum scattering 
albedo with wavelength or an age-dependent extinction are not sufficient to 
explain the previous trend with reddening. Hence, additional extraplanar dust 
has to be invoked. The data suggest that the carriers of the UV bump are 
associated with intermediate-age stellar populations, while they survive in 
the harshest UV-radiation fields owing to dust self-shielding.}
{The existence of different extinction curves implies that different patterns 
of evolution and reprocessing of dust exist at high redshift. Ignoring this 
may produce a non-negligible uncertainty on the star-formation rate estimated
from the rest-frame UV.}


   \maketitle
%

\section{Introduction}\label{introduction}

Extinction occurs whenever electromagnetic radiation propagates through a 
medium containing a mixture of dust grains and molecules. The dependence of 
absorption and scattering by dust from the wavelength of a photon is 
described by the {\it extinction curve}, which depends on the physical and 
chemical properties of dust grains and molecules (Whittet \cite{WHI03}). On a 
galaxy scale, the propagation of electromagnetic radiation through a dusty 
medium is described by the {\it attenuation function}. This is the  
combination of the extinction curve with the geometry of the system, in which 
a substantial fraction of the scattered light is returned to the observer's 
line of sight.

The direct determination of the extinction curve from observations is limited
to our own Galaxy, the Magellanic Clouds, and M\,31 (e.g., Fitzpatrick
\cite{FIT04}; Clayton \cite{CLA04}). The most striking differences among the 
typical extinction curves of the Milky Way (MW) and the Large and Small 
Magellanic Clouds (LMC and SMC, respectively) occur in the UV domain, where 
small dust grains and molecules absorb and scatter with the highest 
efficiency. From the MW to the SMC, the presence of a broad absorption excess
centred at 2175\,\AA{} (the UV bump, see Witt \& Lillie \cite{WIT73} 
and references therein) vanishes (almost) completely, while the slope of the 
extinction curve in the far-UV becomes steeper.

Multiple carriers (i.e., organic carbon and amorphous silicates) were 
suggested to explain the enigmatic, invariant central wavelength and variable 
bandwidth of the 2175\,\AA{} feature (Bradley et al. \cite{BRA05} and 
references therein). The ingredients of interstellar dust that originate the 
UV bump and the far-UV rise are in broadly similar proportions in the MW and 
LMC, but dust seems to be more radically different in the SMC (Whittet
\cite{WHI03}). Differences of the same relevance exist within our own Galaxy
for different sight lines and towards different environments, from dense 
molecular clouds to the diffuse interstellar medium (ISM; see Fitzpatrick 
\cite{FIT04}). They must reflect the extreme sensitivity of the small-sized 
dust components to the local chemical enrichment and energy budget (in terms 
of radiation field and shocks), as well as their selective removal from the 
size distribution owing to a number of physical processes (see Whittet 
\cite{WHI03}).

Nearby starburst galaxies have an extinction curve that seems to lack a 
2175\,\AA{} bump, as the SMC curve, and to have a steep far-UV rise,
intermediate between the MW and SMC curves (Gordon et al. \cite{GOR97}). This 
extinction curve of starbursts was constrained from the modelling of the 
observed UV/optical broad-band colours. An alternative interpretation of the 
empirical attenuation law in starbursts (Calzetti et al. \cite{CAL94}) 
invokes a turbulent interstellar medium (Fischera et al. \cite{FIS03}).
However, it adopts a foreground screen for the distribution of MW dust that is 
reminiscent of the dust/stars configuration of the Gordon et al. 
(\cite{GOR97}) model that reproduces the so-called ``Calzetti law''. This 
configuration is supported by models of large-scale galactic winds driven by 
momentum deposition (Murray et al. \cite{MUR05}).

SMC-like dust seems to be present in Lyman-break galaxies at $z \sim 3$ as 
well (Vijh et al. \cite{VIJ03}). This result is complemented by the finding 
of Noll \& Pierini (2005, \cite{NOL05}), based on spectroscopy instead of 
broad-band photometry. For a sample of 34 massive, UV-luminous galaxies at 
$2 < z < 2.5$, \cite{NOL05} find that the majority of the objects with 
strongly reddened, rest-frame UV spectra have an extinction curve similar to 
the LMC curve. Conversely, the objects with the least reddened, rest-frame 
UV spectra seem to host SMC-like dust. This result strengthens the evidence 
for a difference in the properties of the dusty ISM at intermediate/high 
redshifts arising from previous, sometimes contradictory, results in the 
literature (e.g., Malhotra \cite{MAL97}; Pitman et al. \cite{PIT00}; 
Maiolino et al. \cite{MAI01}; Vernet et al. \cite{VER01}; Hopkins et al. 
\cite{HOP04}; Savaglio \& Fall \cite{SAVF04}; Wang et al. \cite{WAN04}; Wild 
\& Hewett \cite{WIL05}; York et al. \cite{YOR06}).

Here we extend the study of \cite{NOL05} to a heterogeneous sample of 
massive, UV-luminous galaxies at $1 < z < 2.5$. This sample is three times
larger than the \cite{NOL05} one and probes the new redshift range 
$1 < z < 2$. Furthermore, the available ground-based spectroscopy allows us 
to date the oldest stellar populations at least in objects at $1 < z < 1.5$. 
We can also characterise the morphology of a galaxy across the whole redshift 
range in a reliable way from existing imaging with the ACS camera on board 
the Hubble Space Telescope (HST).

Throughout this paper $H_0 = 70\,{\rm km}\,{\rm s}^{-1}\,{\rm Mpc}^{-1}$,
$\Omega_{\Lambda} = 0.7$, and $\Omega_{\rm M} = 0.3$ are adopted. Photometry 
is given in the Vega magnitude system.

\section{The spectroscopic sample}\label{data}

Inferring the presence of the broad dust absorption feature centred at 
2175\,\AA{} either directly (from individual and stacked spectra) or 
indirectly (with the method established in \cite{NOL05}, see 
Sect.~\ref{parameters}) from galaxy spectra taken at optical telescopes
limits the redshift range to $1 < z < 2.5$. The need for rest-frame UV 
spectra with a good continuum definition limits the sample to bright (and 
massive), star-forming galaxies. Finally, the need for a representative 
sample of such galaxies within this redshift range implies selection from 
different spectroscopic surveys.

\begin{figure}
\centering 
\includegraphics[width=8.8cm,clip=true]{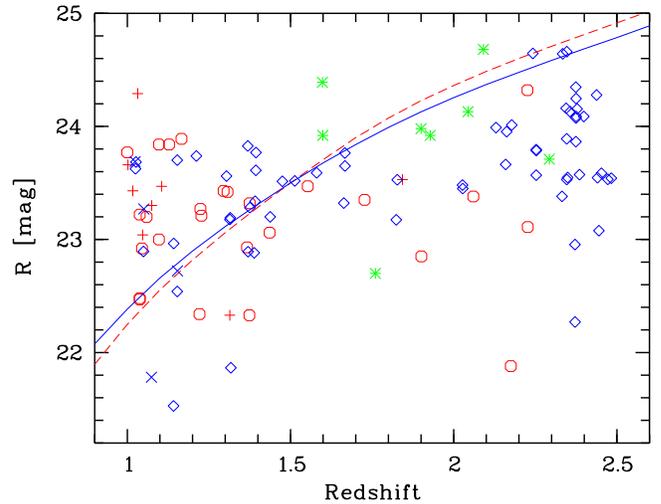}
\caption[]{$R$ magnitude versus redshift for the total sample of 108 actively
star-forming galaxies at $1 < z < 2.5$ selected from the FDF Spectroscopic
Survey (lozenges and crosses), the K20 Survey (circles and plus signs), and 
the GDDS (asterisks). Crosses, plus signs, and asterisks mark galaxies
without a determination of the UV continuum slope. In order to illustrate
selection biases, the curves show lines of constant luminosity for typical
$1 < z < 1.5$ FDF (solid line) and K20 (dashed line) spectra. Both curves
intersect at $R = 23.5$ and $z = 1.5$.}   
\label{fig_R_z}
\end{figure}

\begin{figure}
\centering 
\includegraphics[width=8.8cm,clip=true]{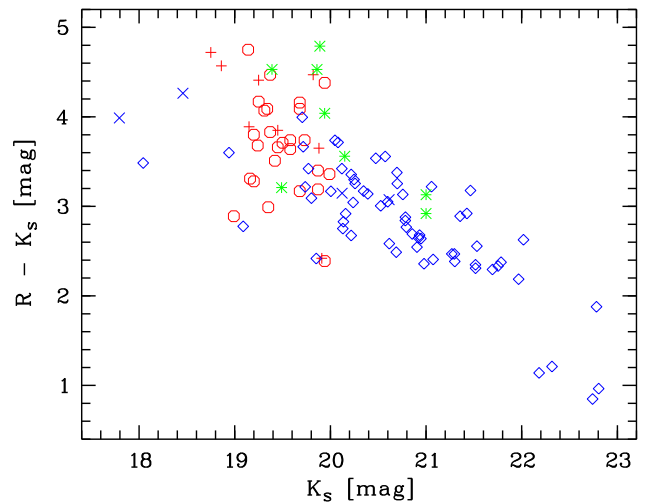}
\caption[]{$R - K_{\rm s}$ versus $K_{\rm s}$ magnitude for the total sample
under investigation. Symbols are the same as in Fig.~\ref{fig_R_z}.}
\label{fig_RK_K}
\end{figure}

For the total sample of 108 massive, star-forming galaxies at $1 < z < 2.5$
selected as described hereafter, Figs.~\ref{fig_R_z} and \ref{fig_RK_K} show
the distribution of the $R$ magnitude (observed frame) as a function of 
redshift and the distribution in the $R - K_{\rm s}$--$K_{\rm s}$
colour--magnitude plane (observed frame), respectively. It is evident that
our sample contains mostly objects with $23 \le R \le 24$ and 
$2 < R - K_{\rm s} < 5$ whatever the redshift, $R - K_{\rm s}$ turning bluer 
towards fainter $K_{\rm s}$ magnitudes. In general, the constraint on the S/N 
of the optical spectra implies that galaxies at $1.5 < z < 2.5$ are more 
luminous in the rest-frame UV than those at $1 < z < 1.5$ (cf. 
Fig.~\ref{fig_L1500_z}), on average. In terms of star-formation activity and 
stellar mass, the total sample is less heterogeneous than what may be 
expected from the different selection criteria of the three surveys of origin 
(cf. Sect.~\ref{globpar_gamma34}). Table~\ref{tab_basic} lists basic 
properties of the sample galaxies. Hereafter we illustrate the different 
selection criteria and properties of the three subsamples.

\subsection{The FDF subsample}\label{FDF}

In addition to the 34 galaxies at $2 < z < 2.5$ with $R < 24.7$ investigated
by \cite{NOL05}, we select 32 objects at $1 < z < 2$ with $R < 24$ from the 
same $I$-limited FORS Deep Field (FDF) Spectroscopic Survey (Noll et al. 
\cite{NOL04}). Only galaxies with a dominant young stellar population (types 
III to V, see Noll et al. \cite{NOL04}) are considered. Basic aspects of the 
data reduction are illustrated in Noll et al. (\cite{NOL04}) and 
\cite{NOL05}.

Ground-based optical/near-IR photometry in nine filters is available from 
Heidt et al. (\cite{HEI03}) and Gabasch et al. (\cite{GAB04}), respectively. 
The FDF was also imaged in the broad-band F814W filter with the HST ACS 
camera. A 10$\sigma$ limit of 26\,mag was reached with four WFC pointings of 
40\,min-exposure each. The data reduction was performed with the standard 
{\sl CALACS\footnote{www.stsci.edu/hst/acs/analysis}} pipeline, and the 
combined final mosaic was produced with the {\sl MultiDrizzle} package
(Mutchler et al. \cite{MUT02}).

\subsection{The K20 subsample}\label{K20}

We select 34 star-forming galaxies at $1 < z < 2.3$ from the spectroscopic 
catalogue of the K20 Survey in the Chandra Deep Field South (CDFS) and a 
field around the quasar 0055-2659 (Cimatti et al. \cite{CIM02}; Mignoli et 
al. \cite{MIG05}). Besides having $K_{\rm s} < 20$, all the selected objects 
but two have $R < 24$. This subsample comprises only K20 objects with a 
secure redshift and a suitable spectral coverage across 2175\,\AA{} 
(basically from 1900 to 2450\,\AA{}). Galaxies with spectral energy 
distributions (SEDs) dominated by old stellar population (class 1 and 1.5, 
see Mignoli et al. \cite{MIG05}) are rejected.

In addition to the basic reduction described in Mignoli et al. 
(\cite{MIG05}), the K20 spectra have been corrected for slit losses using the 
$R$ magnitude of the individual galaxies. Furthermore, they are corrected
for Galactic extinction taking the MW extinction law of Cardelli et al.
(\cite{CAR89}) and the value of Galactic reddening $E(B-V)$ from Schlegel et 
al. (\cite{SCHL98}). Finally, they are mapped to the rest-frame, and smoothed 
to a similar resolution as for the FDF spectra (rest-frame 5 to 8\,\AA{}).

The CDFS was imaged with the HST ACS camera in four different filters as part 
of the ACS GOODS legacy programme (Giavalisco et al. \cite{GIA04}). We make 
use of the F775W band images, where a 10$\sigma$ limit of $26.5$\,mag was 
reached. For the data reduction, we refer the reader to Giavalisco et al.
(\cite{GIA04}).

\subsection{The GDDS subsample}\label{GDDS}

From the $I$ and $K_{\rm s}$-limited Gemini Deep Deep Survey (GDDS, Abraham
et al. \cite{ABR04}), we consider eight additional star-forming galaxies
at $1.5 < z < 2.3$ all of them having $R < 24.7$ and 
$K_{\rm s} \lesssim 21.0$. The limited redshift range is mainly due to the 
relatively narrow wavelength range (typically from 5500 to 9200\,\AA{}) of 
the red GMOS grating used in the GDDS. The GDDS subsample makes half of the 
total sample for $1.5 < z < 2.3$ and consists of galaxies with a prominent, 
young stellar population (class `100', see Abraham et al. \cite{ABR04}).

The basic spectra of Abraham et al. (\cite{ABR04}) have been reduced in the 
same way as described in Sect.~\ref{K20} in order to meet the same standard 
as the spectra of the FDF and K20 subsamples (taken with VLT FORS).

\subsection{The reference sample of nearby starbursts}\label{IUE}

We make use of the same reference sample of 24 local ($z \lesssim 0.02$)
starburst galaxies selected by \cite{NOL05}. For these starbursts,
low-resolution IUE spectroscopy is available for the wavelength range
between 1150 and 3350\,\AA{}.

Sample selection and data reduction are described in \cite{NOL05}.

\section{Data analysis}\label{analysis}

The shape of the extinction curve at rest-frame UV wavelengths is constrained
from a suitable parametric description of the UV SED of a galaxy. This method
was introduced by \cite{NOL05} and is briefly described in 
Sect.~\ref{parameters}. Fundamental properties of a galaxy, like 
star-formation rate (SFR) and stellar mass, are estimated by model fitting 
the optical spectra (Sect.~\ref{sfrmodels} and \ref{massmodels}).
Morphological information comes from model fits of the 2-D surface brightness
distribution of individual galaxies with diffraction-limited HST-ACS imaging
(Sect.~\ref{morph}).

\subsection{A parametric description of the UV continuum}\label{parameters}

\mbox{\cite{NOL05}} introduced a parametrisation of the rest-frame UV SED of 
a galaxy based on five power-law fits to different sub-regions of the UV 
continuum of the same form as in Calzetti et al. (\cite{CAL94}). Narrow 
wavelength regions affected by the presence of strong spectral lines are 
excluded from the fitting procedure for all parameters (cf. Calzetti et al.
\cite{CAL94}; Leitherer et al. \cite{LEI02b}).

Two parameters are particularly suitable to constrain the slope of the 
extinction curve in the far-UV and the presence/absence of the broad
absorption feature centred at 2175\,\AA{}. The first parameter characterises
the apparent strength of the UV bump and is called $\gamma_{34}$. It is
the difference between the continuum slopes measured at 1900 -- 2175\,\AA{}
($\gamma_3$) and 2175 -- 2500\,\AA{} ($\gamma_4$), respectively. A value of 
$\gamma_{34} \sim 1$ indicates the absence of the 2175\,\AA{} feature. This 
is consistent with the presence of SMC-like dust in the ISM of a galaxy. 
Conversely, $\gamma_{34} < -2$ points to an extinction law which exhibits a 
significant UV bump. The second parameter gives the amount of reddening in 
the UV, according to the SHELL models of radiative transfer by Witt \& 
Gordon (\cite{WIT00}). In fact, the relation between amount of dust or 
opacity and reddening is model dependent (e.g., Witt \& Gordon \cite{WIT00}). 
These SHELL models with SMC dust are suitable to describe dust attenuation
in local and high-redshift starbursts (Gordon et al. \cite{GOR97}; Vijh et al.
\cite{VIJ03}). Here the parameter estimating reddening in the UV is defined
as the continuum slope measured at 1750 -- 2600\,\AA{}, with the exclusion of 
the range 1950 -- 2400\,\AA{}. This proxy for the UV reddening by dust is 
called $\beta_{\rm b}$ since it replaces the measure of UV-reddening $\beta$ 
adopted by \cite{NOL05}. In fact, $\beta$ is determined at 1250 -- 1750\,\AA{}
and, thus, is not measurable in optical spectra of objects at $z < 2$.

\begin{figure}
\centering 
\includegraphics[width=8.8cm,clip=true]{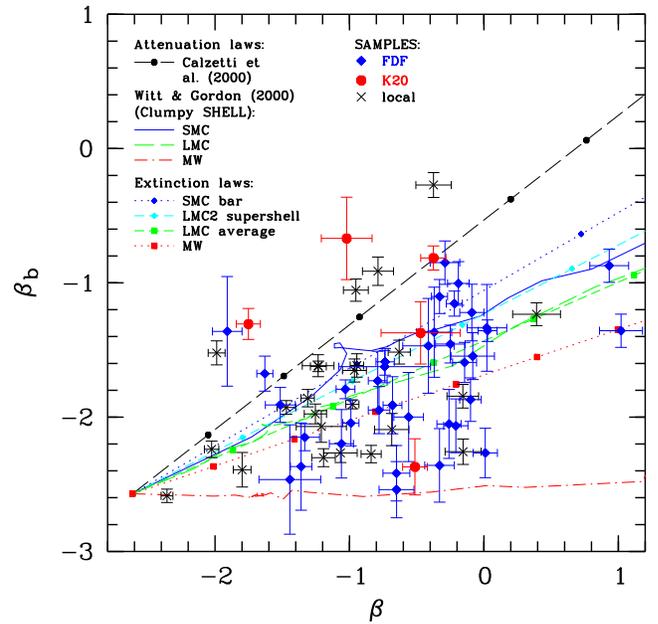}
\caption[]{Comparison of the two UV-continuum slope parameters $\beta$
($1250 < \lambda < 1750$) and $\beta_{\rm b}$ ($1750 < \lambda < 2600$)
obtained from models combining stellar population evolutionary synthesis and 
dust effects as a function of the opacity of the dusty ISM. Dust attenuation 
is described either as for the empirical Calzetti et al. (\cite{CAL00}) law 
(long dashed lines and circles) or as for the Witt \& Gordon (\cite{WIT00}) 
radiative transfer models for the SHELL configuration of dust and stars and a 
two-phase, clumpy ISM with SMC- (solid lines), LMC- (long dashed lines), or 
MW-type dust (dash-dotted lines). Furthermore, we make use of a screen model 
plus the extinction laws of the SMC bar (dotted lines and lozenges), the 
LMC\,2 supershell (dashed lines and lozenges), the LMC average (dashed lines 
and squares), and the MW average (dotted lines and squares) (Cardelli et al. 
\cite{CAR89}; Gordon et al. \cite{GOR03}). Symbols are plotted in intervals 
of $\Delta E(B-V) = 0.1$. In all cases we adopt a Maraston (\cite{MAR05}) 
model with a standard Salpeter IMF, constant SFR, age of 100\,Myr, and solar 
metallicity. The diagram also shows parameter values measured for 24 local 
starburst galaxies (crosses), 33 FDF galaxies at $1.8 < z < 2.5$ (filled 
lozenges), and five K20 objects at $1.9 < z < 2.3$ (filled circles).}
\label{fig_betab_beta}
\end{figure}

In order to establish $\beta_{\rm b}$ as a suitable alternative to $\beta$,
we compare the values of these two parameters determined from synthetic SEDs
(Fig.~\ref{fig_betab_beta}). We make use of different models combining 
stellar population evolutionary synthesis (Maraston \cite{MAR05}) and dust 
attenuation. For data-consistent ages of the stellar populations and 
metallicities (see Sect.~\ref{globpar_gamma34}), the variations in 
$\beta_{\rm b}$ are only of the order of $0.1$. Therefore, in 
Fig.~\ref{fig_betab_beta} we assume the same properties for the stellar 
populations, i.e.: a continuous star-formation over 100\,Myr at a constant 
rate, a standard Salpeter (\cite{SAL55}) initial mass function (IMF), and 
solar metallicity $Z_{\odot}$. Conversely, dust attenuation follows different 
prescriptions (see figure caption). In general, all models show that 
$\beta_{\rm b}$ and $\beta$ increase together when the opacity increases,
with the only exception of models implementing the Witt \& Gordon 
(\cite{WIT00}) SHELL configuration with a two-phase, clumpy medium and 
MW-type dust. In this case, $\beta_{\rm b}$ stays constant ($\sim -2.6$) 
whereas $\beta$ increases when the opacity increases. This behaviour is due 
to the fact that in the spectral regions where $\beta_{\rm b}$ is defined, 
attenuation increases at the same rate with increasing opacity. The 
mapping of $\beta$ into $\beta_{\rm b}$ is model dependent, of course. An 
overall monotonic relation between $\beta_{\rm b}$ and $\beta$ holds even 
when more complex models assuming an age-dependent extinction are considered
(see Sect.~\ref{agedep_extinct}). The locus of these models in the 
$\beta$--$\beta_{\rm b}$ plane only moderately depends on the dust 
distribution, the dust-clearing time scale, and the star-formation history. 
For instance, for composite Witt \& Gordon (\cite{WIT00}) models with 
LMC-type dust and age-dependent extinction we typically find 
$\Delta\beta_{\rm b}$ to be less than $0.2$ for constant $\beta$. In 
conclusion, we can take $\beta_{\rm b}$ as an alternative proxy for the 
UV-continuum reddening, though $\beta_{\rm b}$ has a lower dynamic range
than $\beta$.

Furthermore, the choice of simple but not simplistic radiative transfer 
models to illustrate the relation between $\beta_{\rm b}$ and $\beta$ is 
supported by the fact that most of the data points occupy the region 
delimited by these models in Fig.~\ref{fig_betab_beta}\footnote{We note that 
the distribution of nearby starbursts in Fig.~\ref{fig_betab_beta} shows 
several cases of significant displacement from the locus expected from the 
Calzetti law (Calzetti et al. \cite{CAL94}, \cite{CAL00}). This must not be 
surprising since the Calzetti law is a polynomial fit to the data, and, thus, 
expresses an average behaviour of dust attenuation in nearby starbursts 
across the whole wavelength range of applicability.}. Simultaneous 
measurements of $\beta_{\rm b}$ and $\beta$ were possible for 38 galaxies at 
$1.8 < z < 2.5$ out of the FDF and K20 subsamples, and for the 24 local 
starburst galaxies. Part of the scatter of these data points reflects 
the dependence of $\beta_{\rm b}$ and $\beta$ from the dust properties of 
individual galaxies, as described by the models. Measurement errors in 
$\beta_{\rm b}$ contribute to this scatter as well. They are caused by the 
presence of residuals of night sky lines in the optical spectra obtained for 
the high-redshift galaxies, or by the uncertainties in the correction 
for Galactic extinction and the combination of the spectra of the IUE short- 
and long-wavelength channels for the nearby starbursts (see \cite{NOL05}).

\subsection{Star-formation rates}\label{sfrmodels}

For galaxies at $z > 1.5$, the available optical spectra map rest-frame UV 
wavelengths only. These spectra are contrasted to a suite of synthetic SEDs
across the mapped wavelength domain excluding the range 1950 -- 2400\,\AA{}
and intrinsic (i.e., corrected for attenuation by internal dust) SFRs are 
determined as a best-fit solution. In particular, we build a grid of Maraston 
(\cite{MAR05}) models with constant SFR\footnote{Models with a SFR that 
declines exponentially as a function of time were also explored but they were 
poorly constrained by the data. Therefore they will not be discussed 
hereafter.}, an age between 10 and 1000\,Myr, and metallicity equal to 
0.5, 1, or 2 times solar ($Z_{\odot}$). The ensuing SEDs are attenuated
according to the Calzetti et al. (\cite{CAL00}) law, where the optical colour
excess $E(B-V)$ is the only free parameter. This choice is dictated by
consistency with the literature and simplicity. Hence, the best-fit values of
the intrinsic SFRs have to be taken ``cum grano salis''. In fact, on one 
hand, uncertainty is contributed by the unknown, true attenuation function 
that characterises individual objects, as discussed in Sect.~\ref{results}.
On the other hand, the data only roughly constrain age and metallicity of the 
models. Thus, the degeneracy between the analogous effects of age and dust
reddening on the UV SED is not broken.

For galaxies at $z < 1.5$, the available optical spectra do not map 
wavelengths $\la 1500$\,\AA{}. Therefore, we extrapolate the rest-frame UV 
SEDs of these galaxies down to the Ly$\alpha$ line (1216\,\AA{}) by fitting 
models in the UV domain excluding the range 1950 -- 2400\,\AA{}. To this 
purpose, it is sufficient to adopt a Maraston (\cite{MAR05}) model with age 
of 100\,Myr, solar metallicity, standard Salpeter IMF and constant SFR. The 
ensuing SED is attenuated according to the Calzetti et al. (\cite{CAL00}) law 
as before. The goodness of this procedure is evaluated by comparing the 
UV-continuum slope parameters for fitted synthetic rest-frame UV spectra
and available observed spectra mapping the rest-frame UV down to Ly$\alpha$
(see Sect.~\ref{data}). As a result, these fitted synthetic spectra exhibit
bluer UV continua than the real spectra for 
$\lambda_{\rm min} > 1500$\,\AA{}. Hence, the original fitted values of 
$E(B-V)$ in the Calzetti law were corrected as a function of the measured
value of $\beta_b$ ($\langle \Delta E(B-V) \rangle = 0.04$) and new 
extrapolations were obtained. We estimate that the uncertainty in the 
narrow-band, rest-frame UV luminosity at $1500 \pm 20$\,\AA{}, $L_{1500}$ 
(see Noll et al. \cite{NOL04}), associated with this extrapolation is about 
$0.04$\,dex.

\begin{figure}
\centering 
\includegraphics[width=8.8cm,clip=true]{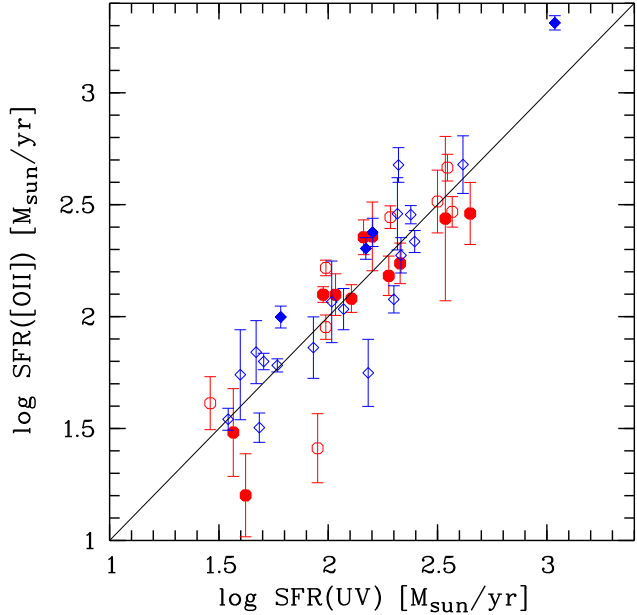}
\caption[]{Comparison of $SFR$\,(UV) derived from model fits to the 
UV/optical spectra and $SFR$\,([O\,II]) derived from [O\,II] fluxes for the 
sample FDF (lozenges) and K20 galaxies (circles) at $1 < z < 1.5$. Objects 
with $\gamma_{34} < -2$ are marked by filled symbols. The $SFR$\,([O\,II]) 
were dust corrected using the Calzetti law as for the $SFR$\,(UV).
The best agreement between these values is obtained for a stars-to-gas
reddening ratio $E(B-V)_{\rm stars} / E(B-V)_{\rm gas} = 0.65$ and is
reproduced here. Errors in the $SFR$\,([O\,II]) only express uncertainties
in the [O\,II] flux.}
\label{fig_compsfr}
\end{figure}

As a consistency check, we estimate SFRs from the [O\,II] emission doublet at 
3727\,\AA{} (e.g., Kennicutt \cite{KEN98}; Kewley et al. \cite{KEW04}) for 
galaxies at $z < 1.5$. We adopt the calibration of Kewley et al. 
(\cite{KEW04}; see their eq.~19), assuming a nearly solar metallicity. We 
make the assumption that the values of the ``instantaneous'' SFR obtained 
from the [O\,II] line emission are statistically the same as the values of 
the SFR averaged over 100\,Myr obtained from the rest-frame UV continuum. 
Fig.~\ref{fig_compsfr} shows that these two sets of estimated SFRs agree, if 
the stellar continuum-to-nebular line-emission reddening ratio is equal to 
$0.65$ ($\sigma = 0.08$), the scatter in the SFR differences being of only 
$0.2$\,dex. This reddening ratio is close to those estimated by Calzetti et 
al. (\cite{CAL00}) and Fernandes et al. (\cite{FER05}), i.e., 
$E(B-V)_{\rm stars} / E(B-V)_{\rm gas} \sim 0.44$ and 0.5, respectively. It 
is significantly less than the value of 1 often assumed in the literature
(e.g., Pierini et al. \cite{PIE05}; Erb et al. \cite{ERB06b}).

\subsection{Stellar masses}\label{massmodels}

Total stellar masses are taken from the literature. They were derived 
following status-of-art techniques that offer consistent results within the 
uncertainties of the photometric data and model assumptions. They are also 
consistent within the uncertainties of a few times $0.1$ dex with the masses 
determined by our SFR estimates (see Sect.~\ref{sfrmodels}). For the FDF and 
K20 subsamples, stellar masses come from Drory et al. (\cite{DRO05}). These 
authors derived the most likely, total stellar masses and their uncertainties
after fitting a large suite of two-component models to the FDF (Heidt et al. 
\cite{HEI03}) and CDFS multi-wavelength photometry (Salvato et al., in 
prep.). Each model consisted of a combination of two composite stellar 
populations: one given by a young (100\,Myr) burst with constant SFR, the 
other by an older ($\ge 500$\,Myr) burst with an exponentially declining SFR. 
Both components of the stellar continuum emission were attenuated 
independently according to Calzetti et al. (\cite{CAL00}). For the GDDS 
subsample, total stellar masses come from Glazebrook et al. (\cite{GLA04}). 
These authors derived the most likely, total stellar masses and their 
uncertainties by evaluating synthetic SEDs against the GDDS photometry of a 
galaxy to determine the mass-to-K-band luminosity ratio, and hence the mass. 
Two-component models plus a standard recipe for attenuation were adopted in a 
way analogous to Fontana et al. (\cite{FON03}) and Brinchmann \& Ellis 
(\cite{BRI00}).

\subsection{Morphological parameters}\label{morph}

The package GIM2D (Simard et al. \cite{SIM99}) is used to fit point-spread 
function (PSF) convolved S\'ersic (\cite{SER68}) profiles to the 
two-dimensional surface brightness of each object, down to a limit of 
$F814W = 24$ (FDF) and $F775W = 24.5$ (CDFS). The S\'ersic model contains
seven free parameters: total intensity, scale-length of the semi-major axis, 
ellipticity $e$, position angle, index $n_{\rm ser}$, and model $x,y$ 
centring. The most valuable parameter is $n_{\rm ser}$, which determines the 
shape of the profile. For each fit, GIM2D automatically determines the 
initial values and limits for the parameter space to be explored using a 
segmentation mask for deblending multiple objects. This mask is provided by 
the standard pipeline for the automated source extraction and photometry 
{\it SExtractor} (Bertin \& Arnouts \cite{BER96}). The PSFs used to convolve 
the model profiles were obtained for each tile by stacking about 10 high S/N 
isolated stars. We put extensive efforts in understanding and quantifying 
how various kinds of systematics (e.g., PSF variations over the field of 
view or dependences on the S/N ratio) could affect the modelling results. To 
this purpose, simulated images with the same characteristics of the true 
ones were produced and analysed in the same way as for the ACS images. GIM2D 
also computes the two indices $R_{\rm T}$ and $R_{\rm A}$ from the thumbnail 
residual image. These indices provide an estimate of the overall smoothness 
of a galaxy image with respect to the fitting model. In other words, they 
estimate the residual substructure like spiral arms in nearby late-type 
galaxies (see the seminal study of Elmegreen et al. \cite{ELM92}), 
peculiarity/asymmetry in the distribution of giant star-forming regions, or 
interaction/mergers in high-redshift galaxies (e.g., Schade et al. 
\cite{SCHA95}).

Furthermore, we characterise the (potentially complex) morphologies of our 
galaxies by means of the model-independent parameters concentration $C$ and 
asymmetry $A$. The $C$-$A$ method was developed in the mid-nineties by 
Abraham et al. (\cite{ABR94}, \cite{ABR96}). Subsequent works (e.g., Wu 
\cite{WU99}; Conselice et al. \cite{CON00}, \cite{CON03}; Menanteau et al. 
\cite{MEN06}) have shown that a better morphological classification is 
obtained by choosing an image pivot point which minimises the measured 
asymmetry. We make use of the CAS parametrisation as proposed and described 
in detail by Conselice et al. (\cite{CON00}, \cite{CON03}). Here it is 
important to say that early-type galaxies have larger concentration and 
lower asymmetry indices than later ones.

\section{Results}\label{results}

Hereafter we show that dust with a UV bump can be present in massive, 
star-forming galaxies at $1 < z < 2.5$ (Sect.~\ref{gamma34_betab}).
Sect.~\ref{gamma34_z} contains reasons why sample selection criteria can 
hinder the inference of the presence of the dust components responsible for
the absorption excess at 2175\,\AA{} (i.e., the carriers of the UV bump) from 
a similar spectral analysis of high-redshift galaxies. We interpret our 
results in Sect.~\ref{agedep_extinct}. Then we investigate how the presence 
of different extinction curves in high-redshift galaxies depends on global 
galaxy properties such as SFR and total stellar mass 
(Sect.~\ref{globpar_gamma34}). Finally, we discuss the effects of topology
and metal enrichment of the ISM on the observed strength of the UV bump
(Sect.~\ref{topology_metals_ISM}).

\subsection{Dust with a pronounced UV bump already exists at high redshift}
\label{gamma34_betab}

\begin{figure}
\centering 
\mbox{\includegraphics[width=7.0cm,clip=true]{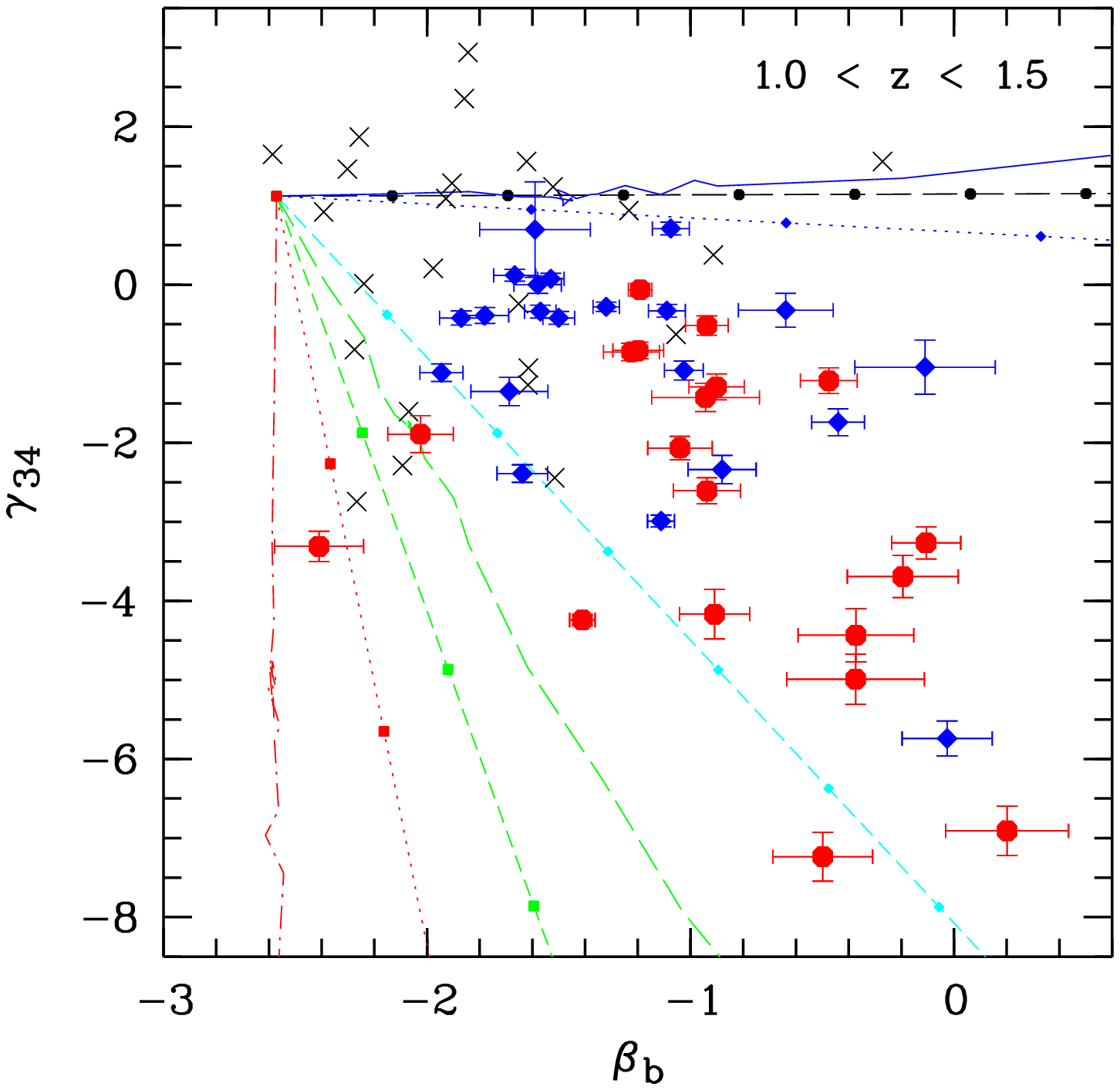}}\\
\mbox{\includegraphics[width=7.0cm,clip=true]{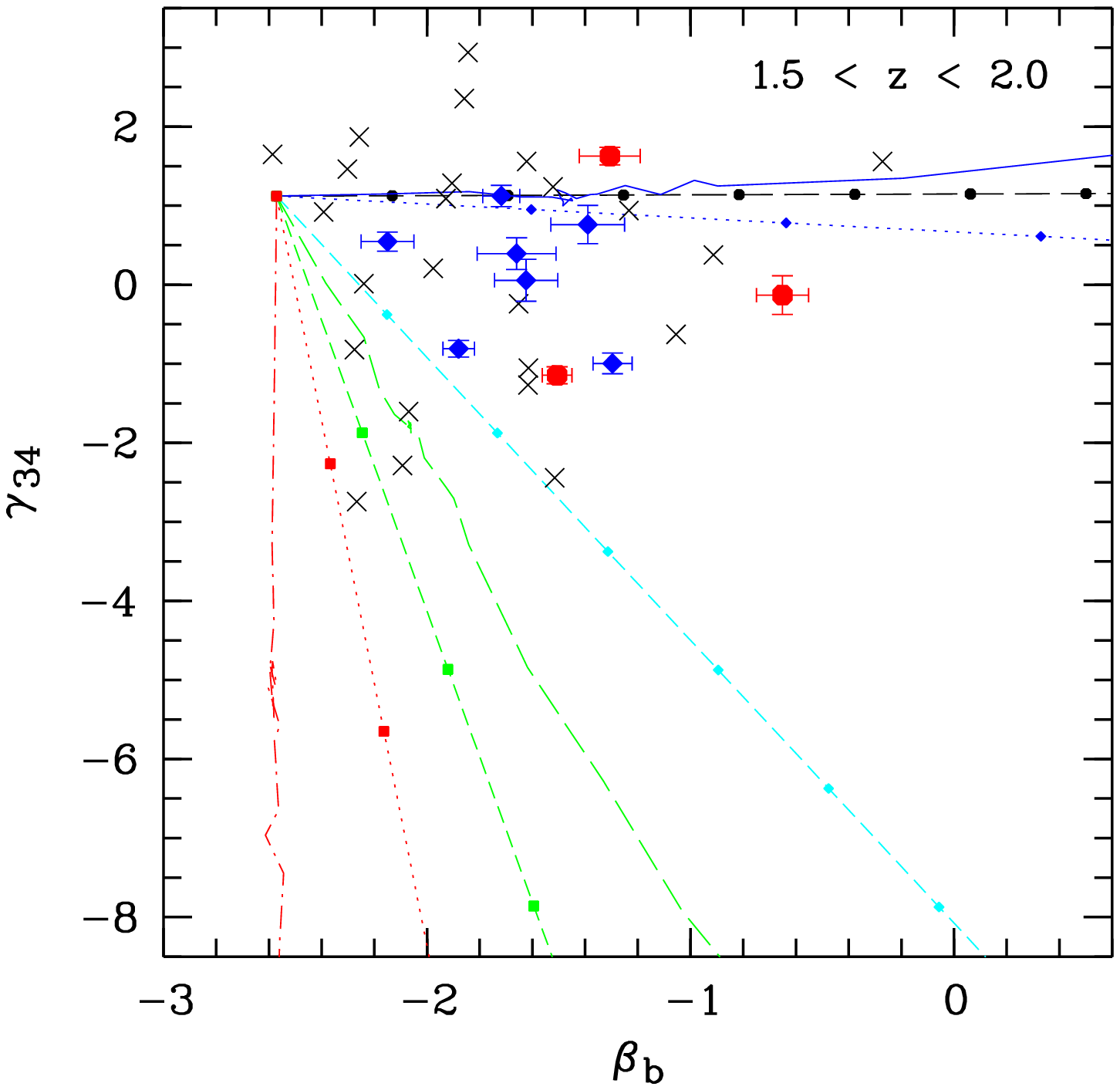}}\\
\mbox{\includegraphics[width=7.0cm,clip=true]{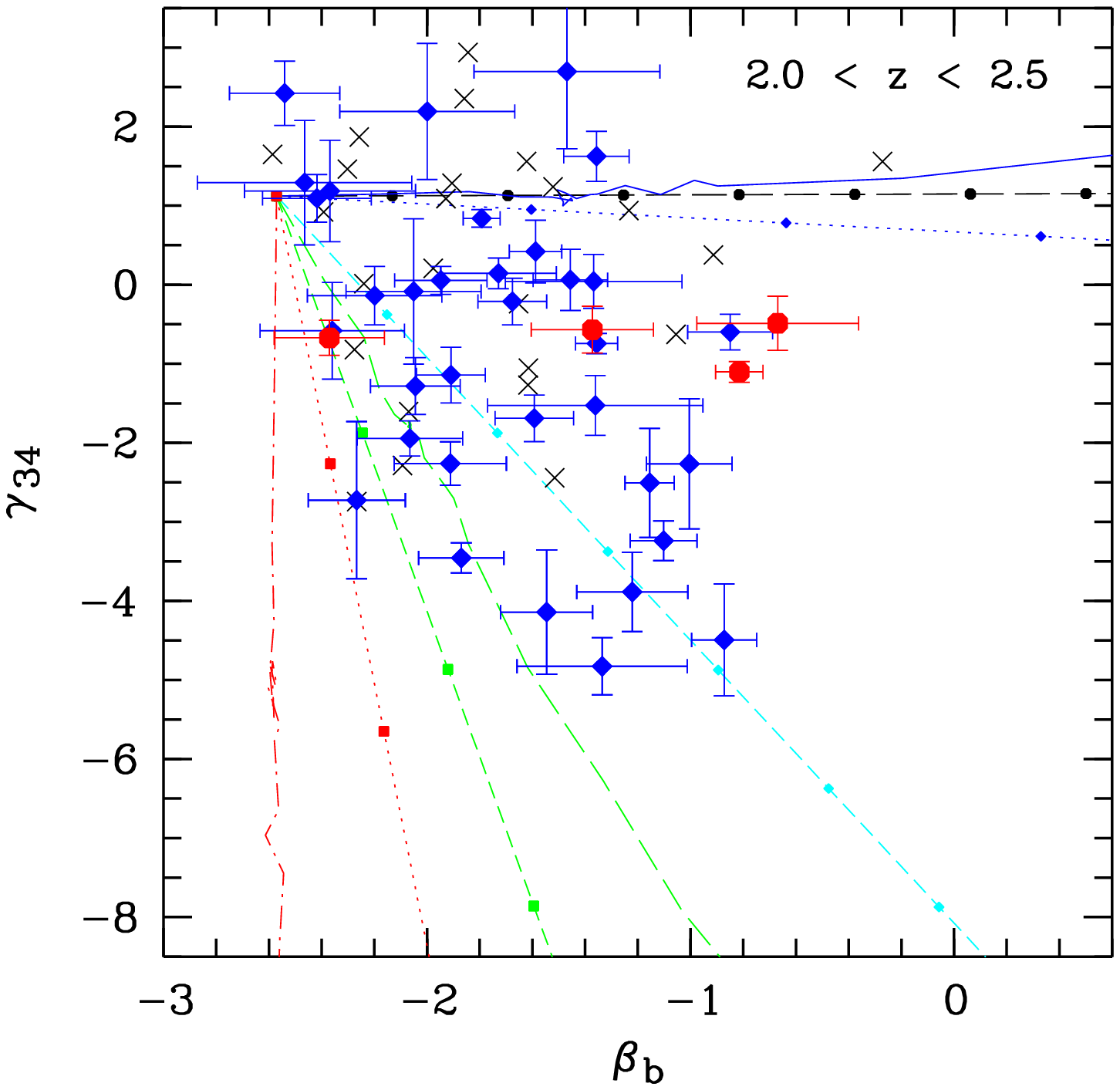}}
\caption[]{Plot of the proxies for the 2175\,\AA{} feature ($\gamma_{34}$) 
and reddening in the UV ($\beta_b$) for the FDF (lozenges) and K20 (circles) 
galaxies listed in Table~\ref{tab_basic} for which both parameters were 
determined. The panels show the distribution of objects at $1 < z < 1.5$ 
(top), $1.5 < z < 2$ (middle), and $2 < z < 2.5$ (bottom), together with that 
of the 24 local starburst galaxies (crosses) as a reference. All panels 
reproduce model tracks obtained for different descriptions of dust 
attenuation as a function of $E(B-V)$ (see Fig.~\ref{fig_betab_beta}). We 
recall that more negative values of $\gamma_{34}$ indicate stronger 
2175\,\AA{} absorption features.}
\label{fig_gamma34_betab}
\end{figure}

\begin{figure}
\centering 
\includegraphics[width=8.8cm,clip=true]{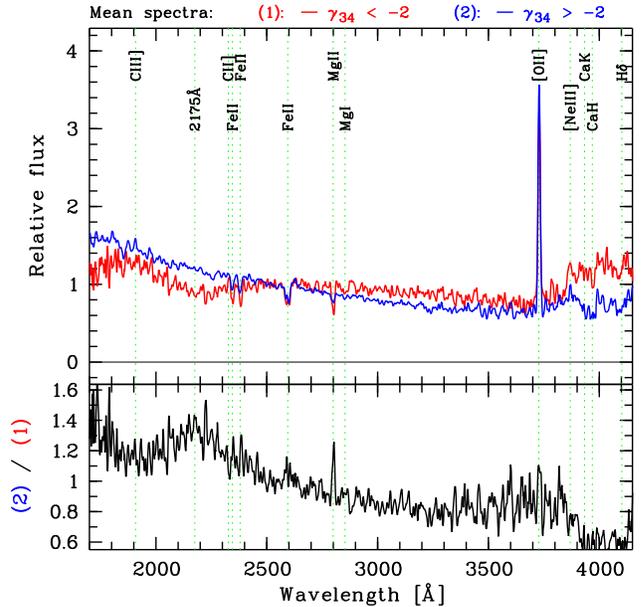}
\caption[]{Comparison of $K_{\rm s}$-weighted composite spectra of 
$1 < z < 1.5$ galaxies with $R < 24$ exhibiting $\gamma_{34} > -2$ (blue or 
dark grey) and $\gamma_{34} < -2$ (red or grey), respectively (top). The 
ratio of both composites, normalised at 2400 -- 2570\,\AA{}, is also shown 
(bottom).}
\label{fig_mean_sb_wb}
\end{figure}

Fig.~\ref{fig_gamma34_betab} shows the proxy for the strength of the 
2175\,\AA{} feature, $\gamma_{34}$, versus the proxy for reddening in the UV, 
$\beta_{\rm b}$, for three different redshift bins, i.e.: $1 < z < 1.5$ (40 
objects), $1.5 < z < 2$ (10 objects), and $2 < z < 2.5$ (38 objects). Tracks 
for the models described in Sect.~\ref{parameters} are also plotted. The 
comparison between the distribution of the observed quantities and the 
expected ones clearly shows that dust with a UV bump and, thus, different 
extinction curves (see Sect.~\ref{agedep_extinct}) do exist at high redshift. 
Now this holds also for $1 < z < 1.5$ (cf. \cite{NOL05}).

Galaxies at $1 < z < 1.5$ exhibit the widest range in the observed strength 
of the UV bump ($-8 < \gamma_{34} < 1$). According to the models described in 
Sect.~\ref{parameters}, dust with a significant UV bump (i.e., 
$\gamma_{34} < -2$) characterises 58\% of the 19 $K_{\rm s}$-selected K20 
galaxies and 19\% of the 21 $I$-selected FDF galaxies at these 
redshifts\footnote{We note that a MW-like extinction law is directly 
suggested only for CDFS-00525.}. This drop is partly caused by selection 
effects (see Sect.~\ref{gamma34_z}). In fact about 29\% of the galaxies in 
the original FDF photometric sample with $1 < z_{\rm phot} < 1.5$ and 
$R < 24$ have $K_{\rm s} < 20$ (Heidt et al. \cite{HEI03}; Gabasch et al. 
\cite{GAB04}). The fraction of FDF galaxies with $\gamma_{34} < -2$ reaches 
33\%, if only the six objects with $K_{\rm s} < 20$ are considered. This 
suggests that dust with a significant UV bump can be present in at least half 
of the $K_{\rm s}$-bright, star-forming galaxies at $1 < z < 1.5$, whatever 
the selection criterion.

Furthermore, Fig.~\ref{fig_gamma34_betab} shows that the 2175\,\AA{} 
absorption feature tends to be associated with larger amounts of reddening at 
UV wavelengths. This confirms the result of \cite{NOL05}. Interestingly the 
largest amounts of reddening appear in the lowest redshift bin. When we 
combine the K20 and FDF objects with $1 < z < 1.5$ to determine the 
$K_{\rm s}$-weighted\footnote{For subsamples the weight of $K_{\rm s}$-bright 
objects differs from the value of 29\% for the whole sample owing to the 
different distributions of the $K_{\rm s}$-bright and $K_{\rm s}$-faint 
galaxies as a function of individual selection quantities.} mean value of 
$\gamma_{34}$, we find that $\langle \gamma_{34} \rangle = -0.44 \pm 0.18$ if 
$\beta_{\rm b} < -1.5$, whereas 
$\langle \gamma_{34} \rangle = -3.48 \pm 0.51$ if $\beta_{\rm b} > -0.5$. 
Consistently, in the same redshift bin, about 80\% of the $K_{\rm s}$-bright 
galaxies with $\beta_{\rm b} > -0.5$ exhibit evidence of dust with a UV bump, 
but only about 40\% of those with $\beta_{\rm b} < -0.5$ do so.

The evidence of a UV bump ($\gamma_{34} < -2$) in 
Fig.~\ref{fig_gamma34_betab} is made clearer by the comparison between the 
$K_{\rm s}$-weighted, composite spectra of galaxies at $1 < z < 1.5$ with 
$R < 24$ and $\gamma_{34} > -2$ or $\gamma_{34} < -2$ 
(Fig.~\ref{fig_mean_sb_wb}). An analogous proof was given by \cite{NOL05} for 
their sample of FDF UV-luminous galaxies at $2 < z < 2.5$ (also reproduced in 
Fig.~\ref{fig_gamma34_betab}). Owing to the different mapping of the 
rest-frame UV domain, the composite spectra in Fig.~\ref{fig_mean_sb_wb} show 
remarkable differences in the strength of the Balmer/4000\,\AA{} break and in 
the Mg\,II doublet at 2800\,\AA{}. The origin of these differences is 
discussed in Sect.~\ref{globpar_gamma34}.

\subsection{Selection effects on the detection of the UV bump}
\label{gamma34_z}

\begin{figure}
\centering 
\includegraphics[width=8.8cm,clip=true]{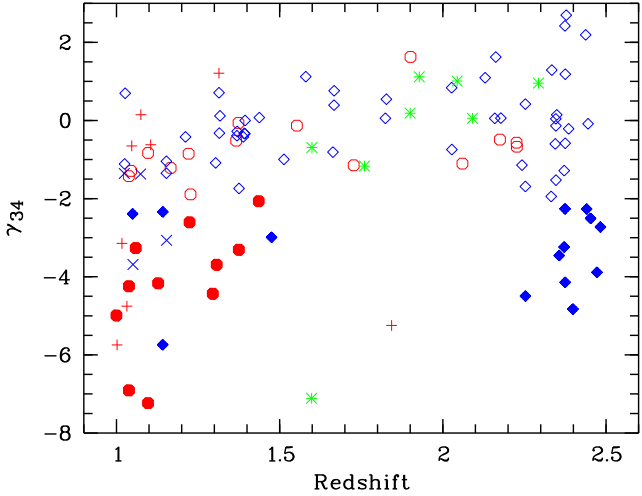}
\caption[]{The distribution of $\gamma_{34}$ values as a function of redshift
for the full sample of 108 massive, star-forming galaxies at $1 < z < 2.5$
selected from the FDF Spectroscopic Survey (lozenges and $\times$), the K20 
Survey (circles and $+$), and the GDDS (asterisks). Crosses and asterisks 
mark galaxies with no possible determination of the UV continuum slope 
$\beta_{\rm b}$ owing to a limited spectral coverage. Filled symbols indicate 
objects with $\gamma_{34} < -2$, i.e., with an extinction curve exhibiting a 
significant UV bump.}
\label{fig_gamma34_z}
\end{figure}

\begin{figure}
\centering 
\includegraphics[width=8.8cm,clip=true]{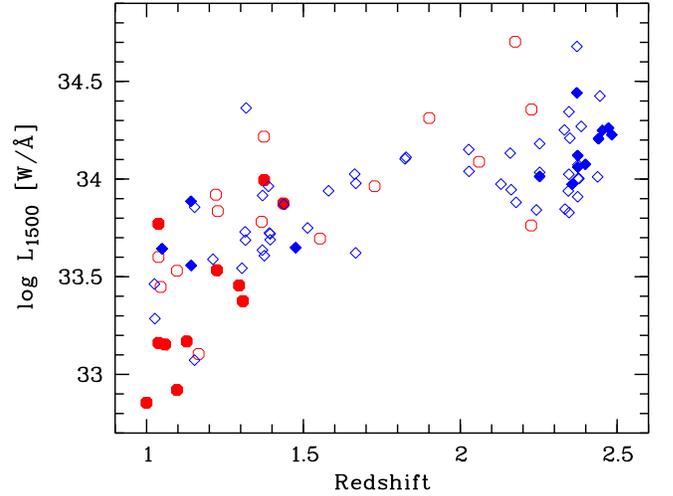}
\caption[]{Luminosity at 1500\,\AA{} versus redshift for 88 galaxies of the 
FDF Spectroscopic Survey (lozenges) and the K20 survey (circles). Galaxies 
with $\gamma_{34} < -2$ are marked by filled symbols.}
\label{fig_L1500_z}
\end{figure}

In Fig.~\ref{fig_gamma34_betab}, the few galaxies at $1.5 < z < 2$ do 
not seem to possess significant 2175\,\AA{} features in their spectra, which 
is at odds with the UV-bump detections at lower and higher redshifts. Hence, 
these objects may not represent the full population of massive, star-forming 
galaxies at $1.5 < z < 2$, though their statistics is poorer than for objects 
in the other two redshift bins. Fig.~\ref{fig_L1500_z} shows that the sample 
galaxies at $1.5 < z < 2.2$ (all exhibiting $\gamma_{34} > -2$) have 
rest-frame luminosities at 1500\,\AA{} that are on average three times higher 
than those of the galaxies with $\gamma_{34} < -2$ at $1.0 < z < 1.5$ but 
slightly less luminous than those with an evident UV bump at $2.3 < z < 2.5$ 
($34.02 \pm 0.06$ versus $34.16 \pm 0.04$ in $\log~L_{1500}$ - in [W/\AA{}], 
respectively). This suggests that our selection can miss galaxies with a low, 
intrinsic UV-luminosity and a significant UV bump, such as those typically 
found at $1.0 < z < 1.5$. However, it could also miss objects with a high, 
intrinsic UV-luminosity and $\gamma_{34} < -2$, such as those found at 
$2.0 < z < 2.5$. This can be understood if the presence of a significant UV 
bump is indeed associated with a larger amount of attenuation at UV 
wavelengths, as suggested by the model interpretation of 
Fig.~\ref{fig_gamma34_betab}. In fact, the sample galaxies at $1.5 < z < 2.2$ 
are mostly selected in $I$ band and exhibit $R < 24$ (cf. Sect.~\ref{data}). 
Now, we note that the $R$ broad-band filter maps exactly the spectral region 
across the UV bump for $1.5 < z < 2.2$. As a check of this hypothesis, we 
plot the distribution of $\gamma_{34}$ as a function of redshift for the full 
sample of 108 massive, star-forming galaxies at $1 < z < 2.5$. With respect 
to Fig.~\ref{fig_gamma34_betab}, Fig.~\ref{fig_gamma34_z} contains additional 
20 galaxies with no measurement of $\beta_{\rm b}$. Among the 26 galaxies at 
$1.5 < z < 2.2$, there are only two with $\gamma_{34} < -2$ against an 
expected number of 8$\pm$3 (corresponding to a fraction of about 30\% as from 
Sect.~\ref{gamma34_betab}). For one (CDFS-00295) the determination of its 
spectroscopic redshift is uncertain. The other (SA15-4762) has $R = 24.4$, 
i.e., it is just fainter than the limiting magnitude ($R = 24$) of the FDF 
and K20 subsamples. We conclude that selection effects can remove from 
the resulting sample very dusty galaxies with dust similar to the LMC or MW 
at $1.5 < z < 2.2$.

\subsection{Extinction curve and dust/stars configuration}
\label{agedep_extinct}

\begin{figure}
\centering 
\includegraphics[width=8.8cm,clip=true]{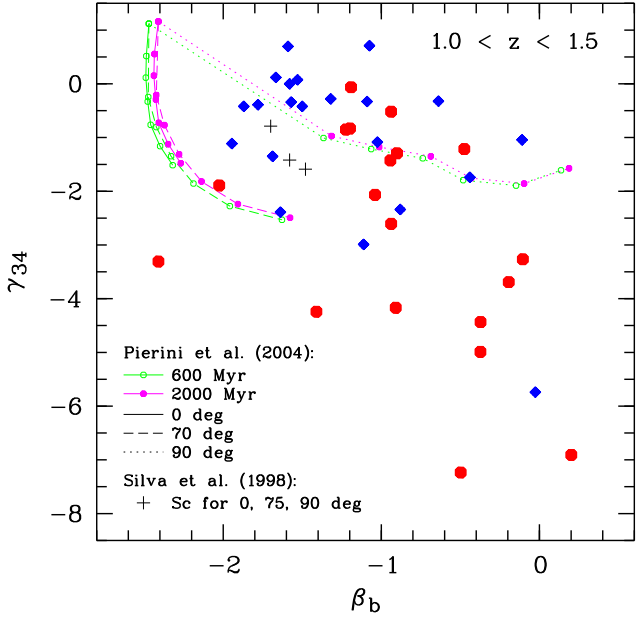}
\caption[]{Comparison among models where stars and dust are distributed
in a disc. First, dust attenuation is described as in Pierini et al. 
(\cite{PIE04}), assuming inclinations of 0 (solid line), 70 (dashed line),
and 90\,deg (dotted line) and continuous star-formation phases of 600
(small open circles) or 2000\,Myr (small filled circles). The $V$-band
opacities plotted are 0, 0.5, 1, 2, 4, 8, and 16. Second, we show Sc-galaxy
GRASIL models (Silva et al. \cite{SIL98}) for inclinations of 0, 75, and
90\,deg (crosses, from the left to the right, respectively).}
\label{fig_dustevol_disc}
\end{figure}

\begin{figure}
\centering 
\includegraphics[width=8.8cm,clip=true]{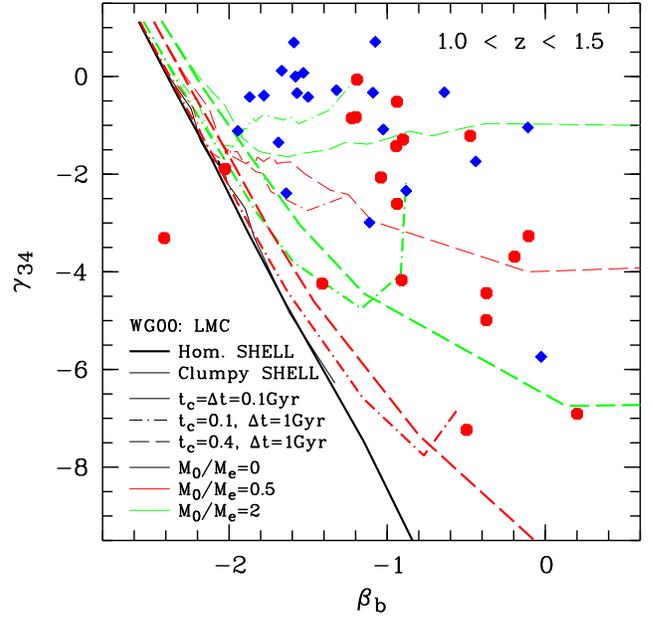}
\caption[]{Comparison between the distribution
in the $\beta_{\rm b}$--$\gamma_{34}$ plane of observed galaxies
at $1 < z < 1.5$ and models where the star-formation activity takes place
in ``macro-regions'' until dust is cleared out there. Dust attenuation
is described as in the Witt \& Gordon (\cite{WIT00}) SHELL configuration,
where the ISM is either two-phase, clumpy (thin lines) or homogeneous
(thick lines) and contains average LMC-type dust. As for $t_{\rm c}$ (see 
Sect.~\ref{agedep_extinct}), it is set equal to either 100\,Myr (dash-dotted 
and solid lines) or 400\,Myr (dashed lines). Star-formation goes on for a 
time equal to the dust clearing time $t_{\rm c}$ (solid, black lines; cf. 
Fig.~\ref{fig_gamma34_betab}) or to 1\,Gyr. For the latter the total stellar 
mass produced by the extinguished stellar population $M_{\rm ext}$ is fixed 
to be half (green or light grey lines) or twice (red or dark grey lines) the 
mass produced by the unextinguished population $M_{\rm unext}$. These 
constraints lead to $SFR_{\rm ext} / SFR_{\rm unext}$ between $0.75$ and 18. 
Three of four star-formation scenarios prescribe the instantaneous SFR at 
$t(z)$ to be above the SFR averaged over the age of the system.}
\label{fig_dustevol}
\end{figure}

In the previous analysis as well as in \cite{NOL05}, it was assumed that dust 
attenuation in massive, UV-luminous galaxies at high redshifts is described 
by the Witt \& Gordon (\cite{WIT00}) Monte Carlo calculations of radiative 
transfer for the SHELL configuration. In particular, a two-phase, clumpy, 
dusty medium is assumed to surround all stars, whatever the ages of the 
stellar populations. If this assumption holds, the interpretation of our 
results is straightforward: different mixtures of dust produce extinction 
curves that are intermediate between the SMC and LMC average extinction 
curves.

However, the previous assumption on the dust/stars configuration can sound
too simple and, thus, be challenged. Therefore, hereafter we discuss two
dust/stars configurations that differ substantially from each other and from
the SHELL configuration in Witt \& Gordon (\cite{WIT00}). In the first case,
we consider a disc geometry together with a different distribution of stars
with respect to the one of the dust according to the different ages of the 
stellar populations. This effect is known as the ``age-dependent extinction'' 
(Silva et al. \cite{SIL98}; Tuffs et al. \cite{TUF04}; Pierini et al. 
\cite{PIE04}; Panuzzo et al. \cite{PAN07}). In the second case, we assume 
that any of our sample galaxies can be ideally pictured as a system where 
stellar populations younger than a certain age (i.e., the dust clearing time 
$t_{\rm c}$) are enshrouded by dust, while older ones live in a dust-free
environment.

We consider two different sets of models describing dust attenuation for a 
disc geometry, namely the ``Sc model'' from the GRASIL library (Silva et al. 
\cite{SIL98}) and the disc models with a two-phase, clumpy, dusty ISM in 
Pierini et al. (\cite{PIE04}). In general, the GRASIL radiative transfer
models describe the fact that new stars are born inside molecular clouds and 
at later times either get rid of their parent environment or move out of it. 
The radiative transfer models described in Pierini et al. (\cite{PIE04})
account for the fact that younger stellar populations are more attenuated by 
dust than older ones by setting the scale-height of the distribution of 
photons equal to lower values the shorter the wavelength. In both cases a 
MW-like extinction curve is assumed. Three values of the inclination of the 
disc are considered in each set of models, namely 0 (i.e., a face-on view),
75 (or 70), and 90 (i.e., an edge-on view) degrees. Furthermore, six different
values of the opacity are considered in the second set of models. In both
cases, intrinsic SEDs are computed for a constant SFR, solar metallicity,
and ages of $0.6$ or 2\,Gyr. As Fig.~\ref{fig_dustevol_disc} shows, the models
do not span a domain as wide as that of the data in
the $\beta_{\rm b}$--$\gamma_{34}$ plane (cf. Fig.~\ref{fig_gamma34_betab}).
In particular, they are confined to $-2.5 \le \gamma_{34} < 1$ although they
assume an extinction curve similar to the average MW one. Differences do exist
however. The GRASIL Sc models span a limited range in $\beta_{\rm b}$ as a
function of the inclination of the disc with respect to the disc models
described in Pierini et al. (\cite{PIE04}), once the opacity is fixed.
Assuming the presence of dust with a UV bump and a smaller scattering albedo
for a shorter wavelength (except for the UV-bump range) would broaden
the range in $\beta_{\rm b}$ spanned by the previous models but would not move
any of them to the region of the highest values of $\beta_{\rm b}$ and
the lowest values of $\gamma_{34}$ (cf. fig.~11 in Inoue et al. \cite{INO06}).
In conclusion, the comparison of observed values and predictions from a more
physical dust/stars configuration for a disc geometry leads us to
the conclusion that especially those galaxies with $\gamma_{34} < -2$ seem
to have dust with a significant UV bump that is partly distributed out of
the disc mid-plane or at least embeds the stellar populations responsible for
the UV emission. This holds whatever the behaviour of the dust scattering
albedo.

Alternatively, we assume that in our high-redshift galaxies new stars are 
born in ``macro-regions'' where the dusty ISM is cleared out after a time
$t_{\rm c}$, which ends the star-formation activity there. For illustrative
purposes only, we make the simplistic case that the stellar populations 
younger than the dust clearing time are screened by dust while the older ones 
do not suffer from extinction at all. As for the value of $t_{\rm c}$, we do 
not take the time scale of the dispersal of the parent molecular cloud by an 
individual supernova, as in Silva et al. (\cite{SIL98}) or in Tuffs et al. 
(\cite{TUF04}), which is of the order of a few Myr (e.g., Leitherer et al. 
\cite{LEI02a}). Conversely, we picture dust clearing as a collective 
phenomenon and set $t_{\rm c}$ equal to 100 or 400\,Myr. Dust attenuation is 
described as for the SHELL configuration in Witt \& Gordon (\cite{WIT00}), 
assuming either a homogeneous or two-phase, clumpy ISM with LMC-type dust. 
Only values of opacity giving a maximum attenuation at 1500\,\AA{} of 5 mag 
are considered. Furthermore, we consider Maraston (\cite{MAR05}) stellar 
population synthesis models with maximum ages of 1\,Gyr and fixed, solar 
metallicity (see Sect.~\ref{stelpop_gamma34}). The total stellar mass 
produced by the extinguished stellar population $M_{\rm ext}$ is fixed to be 
half or twice the mass produced by the unextinguished population 
$M_{\rm unext}$. The SFR is, therefore, changed (divided by a factor from 
$0.75$ to 18) after a time equal to $t_{\rm c}$. Fig.~\ref{fig_dustevol} 
shows a comparison of the previous models and the data in the 
$\beta_{\rm b}$--$\gamma_{34}$ plane, only for galaxies at $1 < z < 1.5$. 
Remarkably, this set of models spans the region defined by the observed 
measurements. For low opacities the light from extinguished, young stars 
dominates the UV spectra and the observed strength of the UV bump increases 
with UV reddening. For higher opacities the contribution of the unobscured, 
older stars becomes more and more important till $\gamma_{34}$ stalls or the 
strength of the 2175\,\AA{} feature even decreases\footnote{An analogous 
decrease of the observed strength of the UV bump (when the fraction of less 
obscured stars older than $t_{\rm c}$ but still emitting UV photons 
increases) was already found by Panuzzo et al. (\cite{PAN07}).}. The latter 
trend is more striking for galaxies with short dust-clearing time scales. The 
most negative $\gamma_{34}$ in our data can only be modelled assuming little 
impact of nearly-unobscured intermediate age stars on the UV SEDs and the 
presence of an almost perfect dust screen for the obscured stellar 
populations. LMC-type dust or dust with even stronger UV bumps are required. 

In conclusion, age-dependent extinction scenarios increase the probability of 
the presence of dust with a significant UV bump in our sample galaxies.

\subsection{Extinction curve and global galaxy properties}
\label{globpar_gamma34}

The carriers of the UV bump exist in galaxies belonging to different
subsamples. 2175\,\AA{} features traced by $\gamma_{34} < -2$ are found for
about 38\% of the FDF galaxies at $2.3 < z < 2.5$. They exist as well in the 
spectra of about 25\% of the $K_{\rm s}$-weighted combined subsample of FDF 
and K20 galaxies with $R < 24$ at $1 < z < 1.5$. This fraction jumps to 
about 52\% if one considers only $K_{\rm s}$-bright galaxies at these 
redshifts (Sect.~\ref{gamma34_betab}). On the other hand, 
Fig.~\ref{fig_L1500_z} shows that galaxies at $1 < z < 1.5$ exhibit lower 
luminosities at 1500\,\AA{} than those at $2 < z < 2.5$, in general. For 
objects with evidence of dust with a UV bump, $L_{1500}$ drops by a factor 
of 4 -- 5. This suggests that the reasons for the presence and/or survival 
of the carriers of the UV bump in high-redshift galaxies are different, at 
least in part, at different redshifts.

Hereafter we investigate if, more generally, the extinction curve depends on 
global characteristics of a high-redshift galaxy like properties of the 
stellar populations, SFR, stellar mass, and morphology. We will consider six
subsamples, which are classified by redshift, UV-continuum slope, and 
observed strength of the UV bump. In particular, we will compare a 
sample at $1 < z < 1.5$ ($K_{\rm s}$-weighted, see 
Sect.~\ref{gamma34_betab}) and one at $2 < z < 2.5$ (only FDF galaxies). 
Both samples are divided into three subsamples each. First, we define
``blue'' galaxies, i.e., galaxies with blue UV continua. For $2 < z < 2.5$,
we take the same classification as introduced by \cite{NOL05}, i.e., 
the ``blue'' galaxies (16 objects) have $\beta < -0.4$. For $1 < z < 1.5$ we 
take $\beta_{\rm b} < -1.5$ (13 objects). The remaining, so-called ``red'' 
galaxies are divided into subsamples for $\gamma_{34} > -2$ (14 objects at 
$2 < z < 2.5$ and 9 ones at $1 < z < 1.5$) and $\gamma_{34} < -2$ (13 and 9 
objects, respectively), since galaxies with relatively strong 2175\,\AA{} 
features are mostly ``red'', i.e., $\beta > -0.4$ and $\beta_{\rm b} > -1.5$, 
respectively. We choose $\beta_{\rm b} = -1.5$ for $1 < z < 1.5$ since 
it is consistent with the discriminating value of $\beta$ for the 
higher-redshift galaxies. In fact, for those objects at $2 < z < 2.5$ with a 
simultaneous measurement of $\beta$ and $\beta_{\rm b}$ we find 
$\beta - \beta_{\rm b} \sim 1$. Comfortably, an analogous difference is 
obtained for the reference sample of nearby starbursts. By chance, 
$\beta_{\rm b} = -1.5$ divides the FDF subsample of galaxies at 
$1 < z < 1.5$ into blue and red subsamples of similar sizes (11 and 10 
objects, respectively).

\subsubsection{Stronger 2175\,\AA{} features in more ``mature'' galaxies}
\label{stelpop_gamma34}

\begin{figure}
\centering 
\includegraphics[width=8.8cm,clip=true]{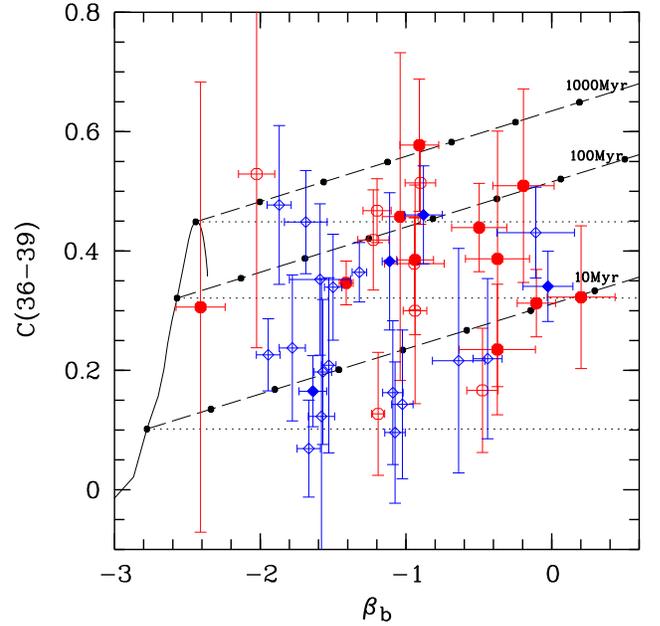}
\caption[]{Balmer-break proxy $C$(36-39) versus reddening measure 
$\beta_{\rm b}$ for the sample FDF (lozenges) and K20 galaxies (circles) with 
$1 < z < 1.5$. Objects with $\gamma_{34} < -2$ are marked by filled symbols. 
Different Maraston (\cite{MAR05}) models for continuous star-formation and 
solar metallicity are shown for estimating the age of the stellar population. 
The dashed lines indicate the effect of continuum reddening by a Calzetti et 
al. (\cite{CAL00}) law on $\beta_{\rm b}$ and $C$(36-39) for ages of 10, 100, 
1000\,Myr. The small filled circles are plotted in intervals of 
$\Delta E(B-V) = 0.1$. Attenuation via the Calzetti law produces
a stronger dependence of $C$(36-39) on $\beta_{\rm b}$ than most other
attenuation laws used in this study (see Fig.~\ref{fig_betab_beta}).
Therefore, the estimated ages would typically increase by a factor of about 2
using SMC and LMC extinction curves. The dust-free cases for the three
considered ages are indicated by the dotted lines.}  
\label{fig_C36-39_betab}
\end{figure}

Fitting the available spectra with the models described in 
Sect.~\ref{sfrmodels} gives ages of less than 100\,Myr (for a constant SFR) 
for the majority of the sample galaxies. This simply confirms the 
star-forming nature of the sample galaxies. Ages of more than 1\,Gyr were 
determined for high-redshift star-forming galaxies like ours by Erb et al. 
(\cite{ERB06b}) and Glazebrook et al. (\cite{GLA04}), based on broad-band 
SEDs mapping the full rest-frame optical domain.

For galaxies at $z < 1.3$, the spectra map also rest-frame optical 
wavelengths, reaching the region of the Balmer/4000\,\AA{} break, where the 
index D4000 is defined (Hamilton \cite{HAM85}; Balogh et al. \cite{BAL99};
Kauffmann et al. \cite{KAU03}). This spectral index is particularly useful to 
constrain ages $\gtrsim 1$\,Gyr, also at high redshifts (e.g., Le Borgne et 
al. \cite{LEB06}). We adopt the definition of D4000 by Balogh et al.
(\cite{BAL99}), where two fluxes (per unit of frequency) are measured in the 
wavelength windows 3850 -- 3950\,\AA{} and 4000 -- 4100\,\AA{}, bracketing 
the Balmer/4000\,\AA{} break. We measure a typical flux ratio of about $1.1$,
which translates into an average age of $50^{+150}_{-35}$\,Myr for Maraston
(\cite{MAR05}) models with constant SFR and attenuated according to the 
Calzetti law, which provides lower age limits (see caption of 
Fig.~\ref{fig_C36-39_betab}). In the complementary dust-free case, this 
estimate increases to $250^{+550}_{-170}$\,Myr. For more complex 
star-formation histories, including both a monotonic behaviour of the SFR as 
a function of time and a random burst of star-formation (cf. Kauffmann et al. 
\cite{KAU03}), a value of D4000 $\sim 1.1$ corresponds to a median age of the 
models equal to 630\,Myr (A. Gallazzi, priv. comm.). The ages of the latter 
models have a large scatter though, ranging from 126\,Myr to 4\,Gyr. True 
characteristic ages will lie somewhere in between the dusty and dust-free 
cases.

Characteristic ages can be determined for all objects at $z < 1.5$, if one 
considers the Balmer-break sensitive pseudo-colour $C$(36-39), defined as the 
flux ratio (in magnitudes) between the fluxes (per unit of frequency) at 3500 
-- 3700\,\AA{} and 3750 -- 3950\,\AA{} (cf. Mignoli et al. \cite{MIG05}). In 
this case, we derive a mean age of $25^{+10}_{-5}$\,Myr from the comparison 
between data and models (Fig.~\ref{fig_C36-39_betab}). If only galaxies at 
$z < 1.3$ are considered, the mean age increases to $80^{+50}_{-30}$\,Myr. 
This is consistent with the indications from D4000. As also suggested by 
Fig.~\ref{fig_mean_sb_wb}, the mean ages of galaxies with $\gamma_{34} < -2$ 
are greater than those of galaxies with $\gamma_{34} > -2$ ($50 \pm 15$\,Myr 
and $20^{+10}_{-5}$\,Myr, respectively). This trend is strengthened in the 
dust-free case, where upper limits to the characteristic age of the stellar 
populations are equal to $380^{+270}_{-160}$\,Myr and $80^{+50}_{-30}$\,Myr 
for galaxies with $\gamma_{34} < -2$ and $\gamma_{34} > -2$, respectively. In 
conclusion, the analysis based on spectroscopic indices reveals the presence 
of intermediate-age ($0.2$ to 1 -- 2\,Gyr-old) stellar populations in 
galaxies at $1 < z < 1.5$. Furthermore, it suggests that galaxies with 
evidence of a UV bump at these redshifts have a larger fraction of 
intermediate-age stars and/or are older than those lacking an evident UV 
bump.

Standard metallicity indicators like the $R_{23}$ index (see, e.g., 
Kobulnicky \& Kewley \cite{KOB04}) are not accessible to optical telescopes
for galaxies at high redshifts. So we estimate the metallicities of galaxies 
at $1 < z < 1.5$ from the mass-metallicity relation evolution model by 
Savaglio et al. (\cite{SAV05}), which was established for galaxies at 
$z < 1$. We obtain $12 + \log({\rm O/H}) = 8.7 \pm 0.2$, i.e., a metallicity 
close to $Z_{\odot}$. We caveat the reader that metallicity estimates based 
on nebular lines overestimate the stellar metallicity by up to 60\% (e.g., 
Bresolin et al. \cite{BRE05}). Nevertheless, a solar-like metallicity is 
supported by a curve-of-growth analysis of interstellar metal absorption 
lines in $1.3 < z < 2$ GDDS galaxy spectra (Savaglio et al. \cite{SAV04}).

For galaxies at $z > 1.5$, we can use the $J - K$ colour as a proxy for the 
characteristic age of a galaxy, since $J - K$ brackets the Balmer/4000\,\AA{} 
break for $2 \leq z \leq 4$. The subsample of FDF galaxies at $2 < z < 2.5$ 
exhibits $\langle J - K \rangle = 1.7$ ($\sigma_{J-K} = 0.4$). Hence these 
galaxies do no meet the $J - K > 2.3$ criterion for distant red galaxies at 
$z > 2$ (DRGs, Franx et al. \cite{FRA03}; van Dokkum et al. \cite{DOK04}), so 
that the bulk of their stellar populations is most probably not much older 
than 1\,Gyr\footnote{This age constraint is compatible with the best-fit age 
estimates of three DRGs ($0.3$, $1.3$, and $1.4$\,Gyr) of Kriek et al. 
(\cite{KRI06}), based on D4000 indices ranging from $1.2$ to $1.4$.} (see 
also Pierini et al. \cite{PIE05}). A certain degree of evolution is required 
by the level of enrichment of the stellar populations reached by galaxies at 
$2 < z < 2.5$ in the FDF subsample (see \cite{NOL05}). In fact, for all FDF 
galaxies at these redshifts with available high-S/N spectroscopy at 
rest-frame UV wavelengths, Mehlert et al. (\cite{MEH02}, \cite{MEH06}) 
determined typical metallicities of 0.5 -- 1\,Z$_{\odot}$, using the 
stellar-wind feature C\,IV\,$\lambda\,1550$ and the photometric indices 
`1370' and `1425' (Leitherer et al. \cite{LEI01}), respectively. This range 
is in agreement with the mass-metallicity relation for $z \sim 2$ of Erb et 
al. (\cite{ERB06a}).

\subsubsection{No trend between extinction curve and star-formation rate or 
stellar mass}\label{SFR_mstar_gamma34}

Table~\ref{tab_avpar} lists average values of the best-fit SFR obtained for 
individual galaxies (Sect.~\ref{sfrmodels}) that are grouped in different
subsamples. In general, less reddened galaxies exhibit lower SFRs at any
redshift as a consequence of a lower value of the best-fit amount of
attenuation. In fact, blue subsamples typically exhibit $E(B-V) \sim 0.2$ -- 
$0.3$, while red subsamples seem to have $E(B-V) \sim 0.4$ on average, 
whatever the redshift (Table~\ref{tab_avpar}).

We note that the SFRs of the $2 < z < 2.5$ subsamples are generally higher 
by a factor of 2 to 4 than those of the corresponding $1 < z < 1.5$ 
subsamples. The largest differences are observed for red galaxies associated 
with a significant UV bump. In fact, red galaxies at $2 < z < 2.5$ tend to 
exhibit a stronger absorption feature centred at 2175\,\AA{} the larger the 
attenuation, and, thus, the higher the SFR. Consistently, these galaxies are 
much redder than typical Lyman-break galaxies (cf. Shapley et al. 
\cite{SHA04}; Noll et al. \cite{NOL04}) and have higher SFRs (about 
500\,M$_{\odot}$ yr$^{-1}$ on average). Their SFRs exceed those estimated 
for galaxies selected by UV colour indices (e.g., Erb et al. \cite{ERB06b}) 
by about one order of magnitude. Daddi et al. (\cite{DAD04}) estimated SFRs 
of almost the same order of magnitude for the K20 galaxies at the highest 
redshifts in our sample.

\begin{figure}
\centering
\mbox{\includegraphics[width=8.8cm,clip=true]{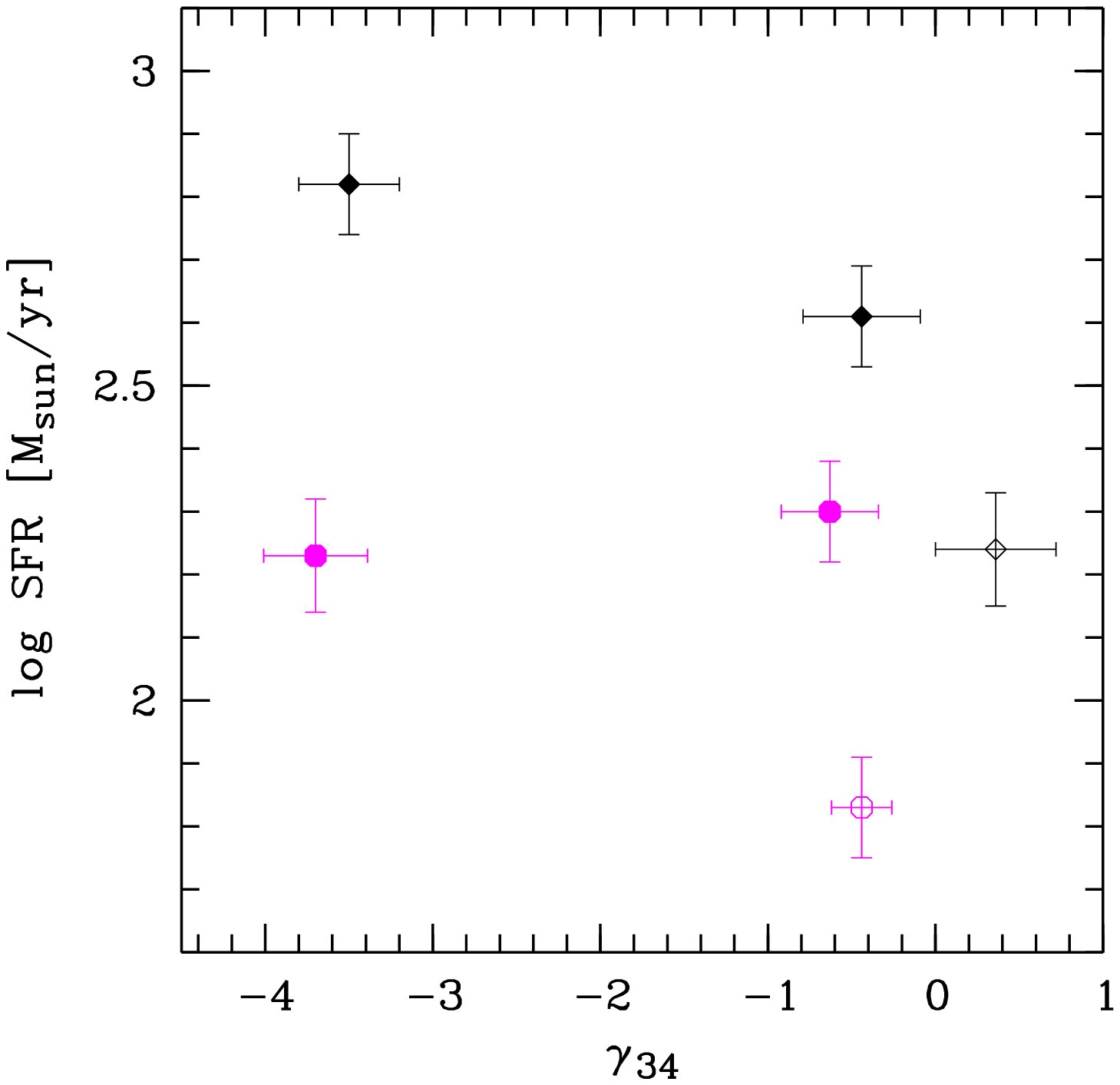}}\\
\mbox{\includegraphics[width=8.8cm,clip=true]{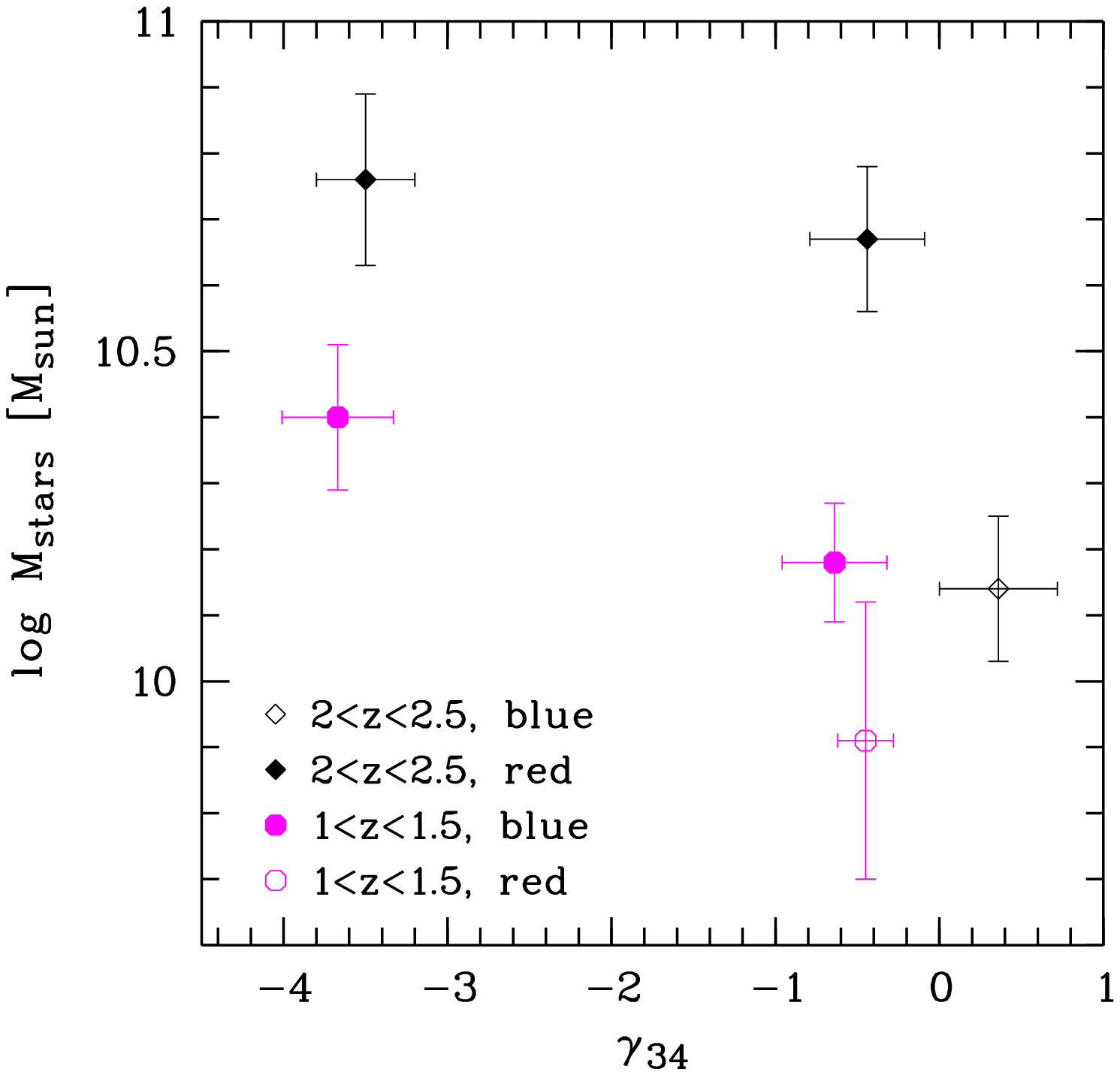}}
\caption[]{SFR (top) and total stellar mass (bottom) versus the proxy for the 
2175\,\AA{} feature $\gamma_{34}$ for the six subsamples defined in 
Sect.~\ref{globpar_gamma34}. Lozenges and circles indicate the FDF 
$2 < z < 2.5$ and the FDF+K20 $1 < z < 1.5$ galaxies, respectively. Open 
symbols mark the subsamples with low reddening ($\beta < -0.4$; 
$\beta_{\rm b} < -1.5$), while the highly reddened galaxies are represented 
by filled symbols. Mean errors are indicated.}
\label{fig_baspar1_gamma34}
\end{figure}

In addition, Fig.~\ref{fig_baspar1_gamma34} shows the average total stellar
mass as a function of $\gamma_{34}$ for different subsamples. The average 
total stellar mass is equal to $1.3 (\pm 0.3) \times 10^{10}$\,M$_{\odot}$
and $2.8 (\pm 0.5) \times 10^{10}$\,M$_{\odot}$ for galaxies at $1 < z < 1.5$
and $2 < z < 2.5$, respectively. It appears to be larger in star-forming
galaxies at $1 < z < 2.5$ that suffer from a larger amount of UV-continuum
reddening. However, there is no trend between observed strength of the UV 
bump and total stellar mass, as already found by \cite{NOL05}.

Finally, we determine specific SFRs, i.e., SFRs per unit of stellar mass.
As shown in Table~\ref{tab_avpar}, the galaxies in our sample exhibit a 
narrow range in specific SFR ($\phi \sim 10^1$\,Gyr$^{-1}$) whatever the 
redshift. This is a rather large value, especially when considering that
$\phi$ has a range of four orders of magnitude for much larger samples that 
probe the mixture of galaxies at high redshift (e.g., Reddy et al.
\cite{RED06}). A plausible reason for the narrow range of high $\phi$ of our 
galaxies is that they are all close to the peak of their star-formation 
history.

\subsubsection{Extinction curve and morphology}\label{morph_gamma34}

\begin{figure}
\centering
\mbox{\includegraphics[width=8.8cm,clip=true]{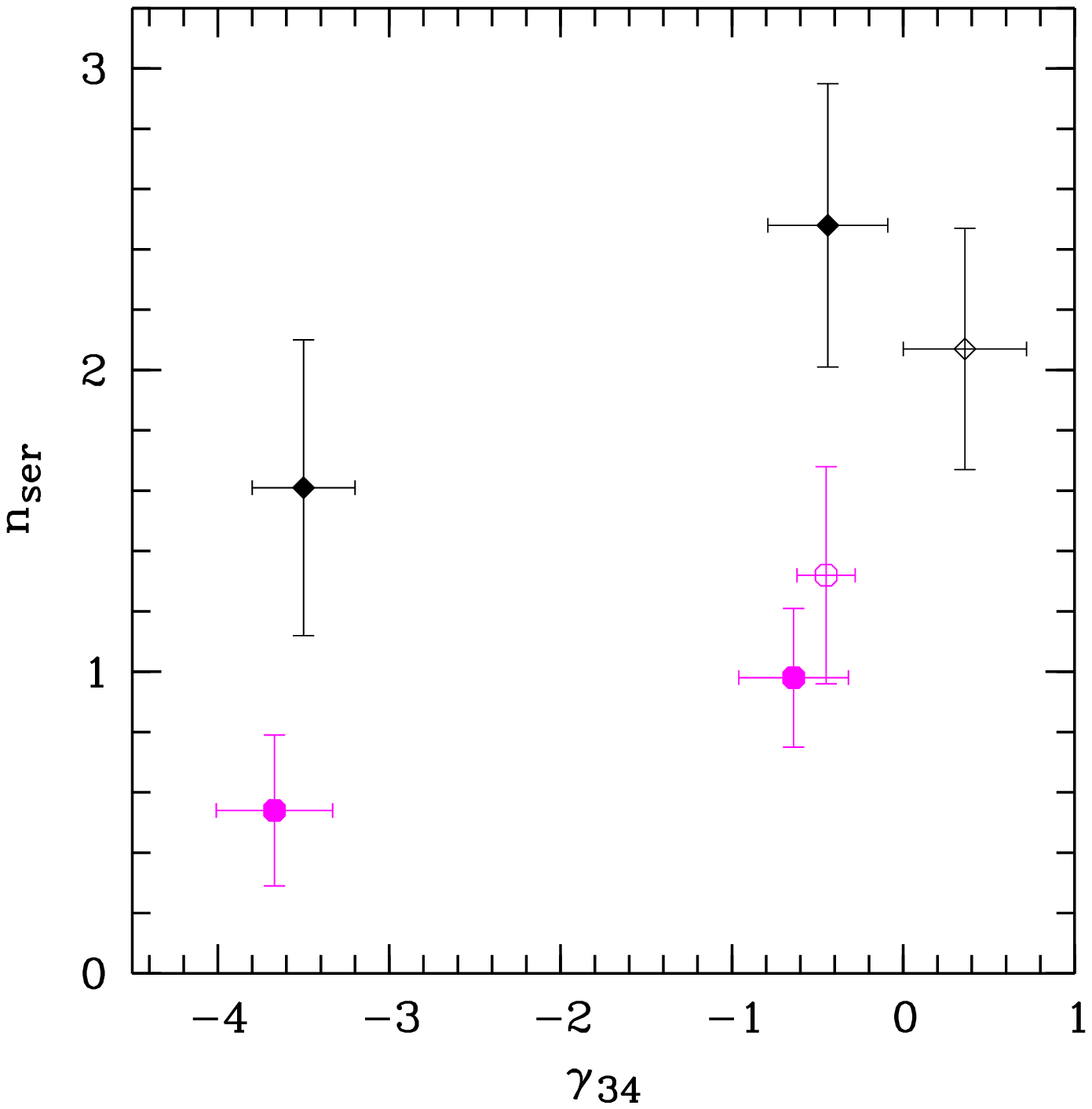}}\\
\mbox{\includegraphics[width=8.8cm,clip=true]{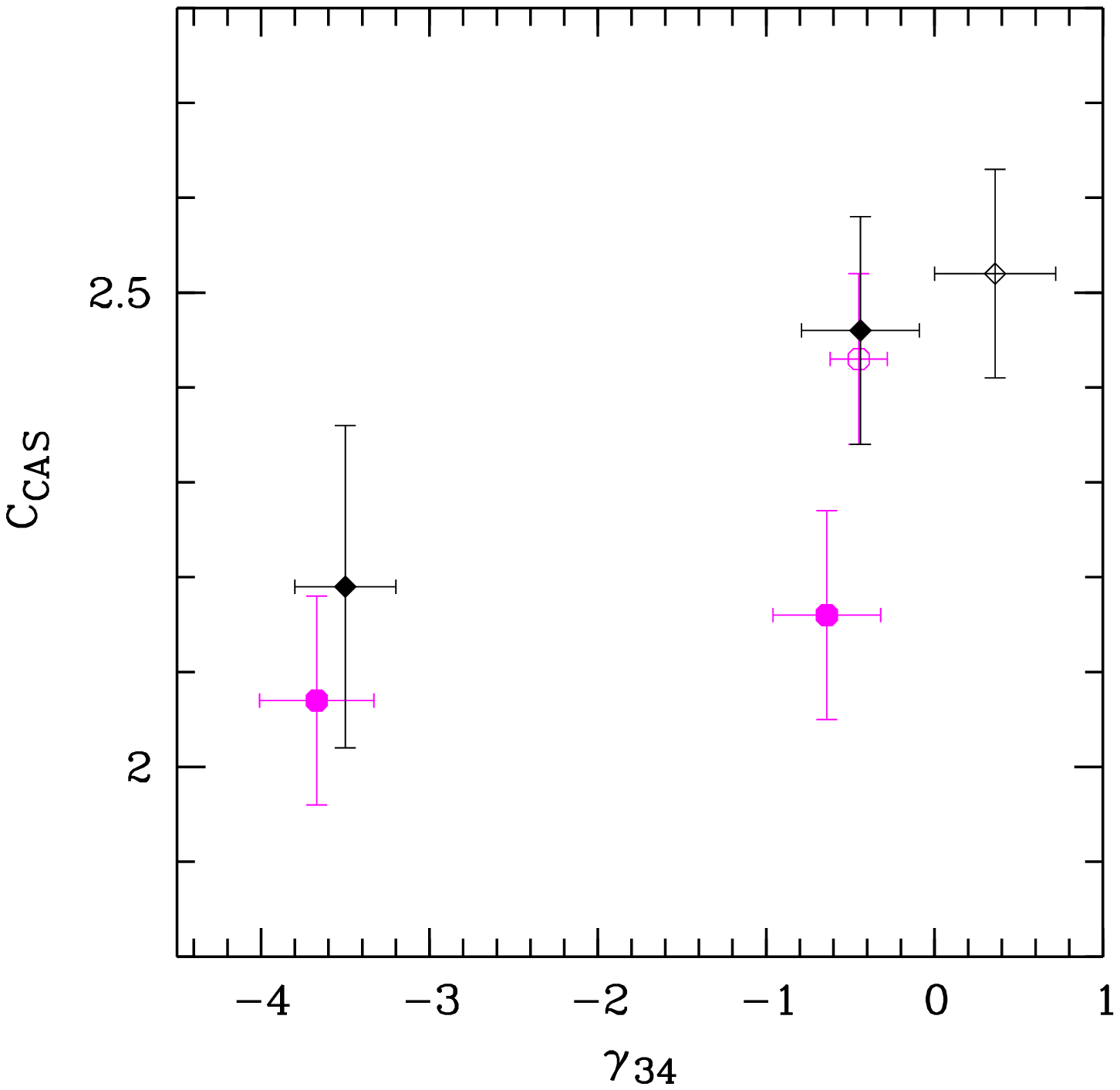}}
\caption[]{S\'ersic index (top) and concentration (bottom) versus the proxy
for the 2175\,\AA{} feature $\gamma_{34}$ for the six subsamples defined in 
Sect.~\ref{globpar_gamma34}. Symbols are as in 
Fig.~\ref{fig_baspar1_gamma34}. Mean errors are indicated.}
\label{fig_baspar2_gamma34}
\end{figure}

\begin{figure}
\centering 
\includegraphics[width=8.8cm,clip=true]{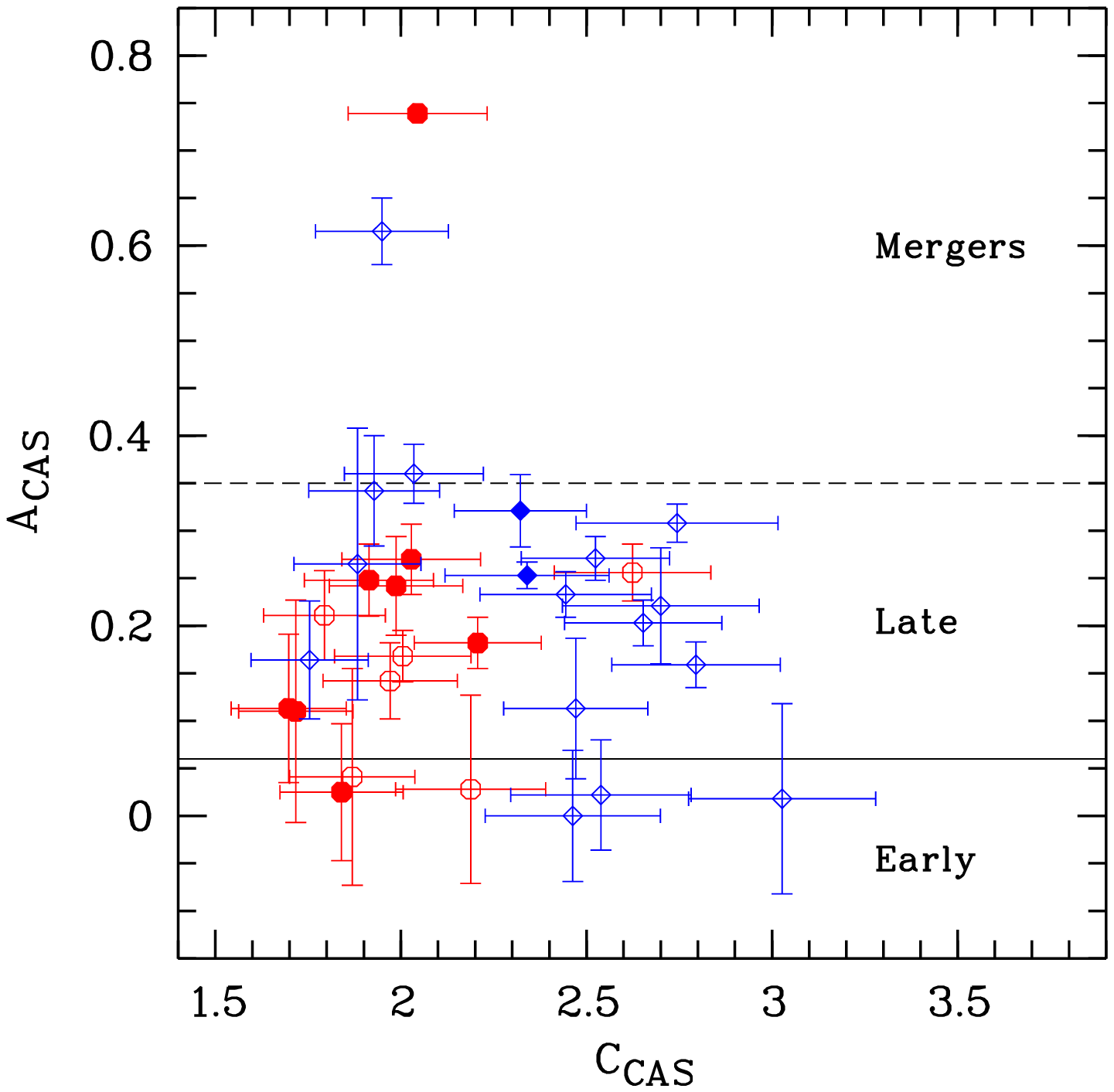}
\caption[]{CAS parameters concentration $C$ and asymmetry $A$ (Conselice et 
al. \cite{CON00}, \cite{CON03}) for our sample of FDF (lozenges) and K20 
galaxies (circles) at $1 < z < 1.5$. Galaxies with $\gamma_{34} < -2$ are 
marked by filled symbols. The solid and dashed lines separate early-type,
late-type, and merging galaxies by their asymmetry (see Conselice et al. 
\cite{CON03}).}
\label{fig_A_C}
\end{figure}

Table~\ref{tab_avpar} lists average values of the effective radius 
$R_{\rm e}$, S\'ersic index $n_{\rm ser}$, concentration $C$, asymmetry $A$, 
and ``clumpiness'' $R_{\rm T} + R_{\rm A}$ of galaxies grouped in the six
subsamples that were described previously. We stress that these morphological
parameters were determined in different rest-frame wavelength domains from 
the mid-UV to the $U$ band depending on redshift (see Sect.~\ref{morph}). In 
general, the rest-frame UV/$U$-band morphology of a galaxy does not appear to 
be directly related with the shape of the extinction curve at UV wavelengths 
whatever the redshift. Nevertheless, blue objects at $2 < z < 2.5$ and 
$1 < z < 1.5$ tend to have smaller effective radii than red objects at the 
same redshift, whether or not a UV bump is detected in their spectra (see 
also \cite{NOL05}).

As for the S\'ersic index, there is a hint that red galaxies with evidence of 
a UV bump have a lower $n_{\rm ser}$ than red galaxies with no evidence of it 
in both redshift bins (see Fig.~\ref{fig_baspar2_gamma34}). For instance, at 
$1 < z < 1.5$, the former exhibit an average S\'ersic index of 
$0.54 \pm 0.25$, the latter one equal to $0.98 \pm 0.23$. Overall, the 
S\'ersic index seems to be larger for galaxies at higher redshifts 
(Fig.~\ref{fig_baspar2_gamma34}). Since the available high-resolution imaging
probes the rest-frame UV/$U$-band morphology, the S\'ersic profile describes
the large-scale distribution of star-formation regions instead of the 
classical Hubble type. Hence, the azimuthally-averaged radial distribution of 
the rest-frame UV/$U$-band light traced by $n_{\rm ser}$ is shallower for 
galaxies with dust producing a significant 2175\,\AA{} feature with respect 
to those without it, whatever the redshift. Furthermore, it is more peaked at 
higher redshifts than at lower ones, despite this may be an effect of the 
cosmological brightness dimming. In fact, a faint component of a galaxy like 
a disc can fail detection more easily at higher redshifts, thus producing a 
spuriously larger value of $n_{\rm ser}$.

In addition to the relatively large effective radii and low S\'ersic indices,
the low values of concentration of the red galaxies with evidence of a UV 
bump at $1 < z < 1.5$ (see Fig.~\ref{fig_baspar2_gamma34}) suggest that most
of these galaxies are large systems with shallow radial profiles at 
rest-frame UV/$U$-band wavelengths. This is not surprising since most of the 
galaxies at these redshifts can be classified as late types (see 
Fig.~\ref{fig_A_C}). However, the visual inspection of the ACS images reveals 
that the fraction of objects with a shallow light profile in the rest-frame 
UV/$U$-band is 40 -- 70\% if $\beta_{\rm b} > -1.5$, but only 20 -- 40\% if 
$\beta_{\rm b} < -1.5$. For red galaxies with a significant UV bump, this 
fraction becomes larger (60 -- 80\%), though there are exceptions: CDFS-0271, 
the object with the strongest observed 2175\,\AA{} absorption feature, 
appears as a quite compact galaxy ($R_{\rm e} = 2.5$\,kpc).

Establishing the rest-frame UV morphology of a galaxy at $2 < z < 2.5$ is not 
possible from the available data, owing to the increased cosmological 
dimming. Nevertheless, we can investigate the presence of strongly distorted
morphologies and/or multiple main components. These characteristics exclude
the possibility of a single object with a smooth, radial surface brightness
profile in the rest-frame UV. While at $1 < z < 1.5$ almost all galaxies show
only one major component, at $2 < z < 2.5$ 45 -- 65\% of the objects seem to 
have two or more main components. There is not much difference between red
($\beta > -0.4$) and blue galaxies ($\beta < -0.4$) as for the fraction of 
objects with multiple components. However, there is a considerable difference 
when only red objects with strong 2175\,\AA{} features are considered. For 
these galaxies, the fraction of objects with multiple components rises to 70 
-- 80\%. This suggests that galaxies with extinction curves exhibiting a 
significant UV bump at $2 < z < 2.5$ are either systems with many, large 
star-formation complexes or merging systems. This could point to the 
existence of an intrinsic structural difference with respect to analogous 
galaxies at $1 < z < 1.5$, which appear as smooth, disc-like systems in the 
rest-frame UV/$U$-band.

\subsection{Properties of the dusty interstellar medium}
\label{topology_metals_ISM}

For the subsample of galaxies at $2 < z < 2.5$, the strength of the 
2175\,\AA{} feature is linked to the average equivalent width (EW) of six 
prominent, interstellar low-ionisation, absorption lines $W_{\rm LIS}$ 
mapped by the FDF spectra (\cite{NOL05}). This relation suggests that the 
presence of a significant UV bump is linked to high column densities of 
(cold) gas and dust in the direction towards the observer, since 
$W_{\rm LIS}$ traces the covering factor of neutral-gas clouds (Shapley et 
al. \cite{SHA03}). In this scenario, the carriers of the UV bump are 
protected from strong and hard radiation fields by other (more robust) dust 
grains (e.g., Gordon et al. \cite{GOR03}). For objects at $1 < z < 1.5$, 
there is a lack of strong, purely interstellar absorption lines at the 
rest-frame UV wavelengths mapped by optical telescopes. Nevertheless, one 
can investigate the most striking absorption features, i.e., the Fe\,II 
doublet at 2600\,\AA{} and the Mg\,II doublet at 2800\,\AA{}. We find that 
the strength of the Fe\,II absorption does not depend on $\beta_{\rm b}$ or 
$\gamma_{34}$, being $W_{\rm Fe\,II} \sim 5$\,\AA{} for all subsamples. On 
the other hand, $W_{\rm Mg\,II} = 1.8 \pm 0.7$\,\AA{} for blue galaxies but 
rises to $5.0 \pm 0.5$\,\AA{} for red galaxies with a strong 2175\,\AA{} 
feature. Hence, the strength of the Mg\,II absorption increases together with 
the apparent strength of the UV bump. The existence of this trend is robust 
against the contamination of the measured Mg\,II equivalent width by an 
emission contribution from an active galactic nucleus (AGN) since the 
presence of AGN in the sample is excluded by the line profiles.

The explanation of these results is not straightforward, given the multiple
origin of the absorption in the Fe\,II and particularly the Mg\,II doublets.
In fact, the latter feature originates in stellar atmospheres as well as in 
the ISM and, thus, depends on the fraction of intermediate-type stars
(particularly A stars), the metallicity and chemical composition of the 
stellar populations as well as of the ISM, and the structure of the ISM. With 
these caveats in mind, even a significant contribution of photospheric 
absorption (see, e.g., Fanelli et al. \cite{FAN92}) to the strength of MgII 
will not harm our viable interpretations, that are based on a purely 
interstellar origin of the MgII and FeII lines. First of all, we note that Mg
and Fe have similar condensation temperatures and their depletions exhibit
a similar systematic trend with density (Whittet \cite{WHI03}). However their
relative gas-phase abundances depend on the gas temperature (Savage \& 
Sembach \cite{SAV96}). In the warm- and cold-gas phases, there is less than 
10\% and 1\% of Fe, respectively. Furthermore, Fe is depleted on dust grains 
about six times more than Mg in the cool ISM, but the Mg-to-Fe gas-phase 
abundance ratio decreases to about 2 in the warm ISM (Savage \& Sembach 
\cite{SAV96}). Hence, a larger Mg-to-Fe equivalent width ratio can be 
explained independently by a greater metallicity (plus depletion on dust 
grains), as a larger fraction of the ISM being cool or as the ISM being 
particularly enriched in Mg. Sources of enrichment of the ISM, with a high 
yield of Mg but only a modest one (if any) of Fe, can be identified as the 
Asymptotic Giant Branch (AGB) stars. They dominate the overall mass-loss 
budget at least in the Galaxy (Whittet \cite{WHI03}). Both the presence of 
these intermediate-mass (2 -- 5\,M$_{\odot}$) stars and a higher metallicity 
are naturally consistent with the ages estimated for red galaxies with 
evidence of a UV bump and a larger Mg-to-Fe equivalent width ratio at 
$1 < z < 1.5$ (see Sect.~\ref{stelpop_gamma34}).

\section{Discussion}\label{discussion}

In the local Universe, well-characterised extinction curves exist for 
different environments in the Milky Way and the Magellanic Clouds. From the 
MW to the SMC, the presence of a broad absorption feature centred at 
2175\,\AA{} (the UV bump, see Witt \& Lillie \cite{WIT73} and references 
therein) vanishes (almost) completely, while the slope of the extinction 
curve in the far-UV becomes steeper. These differences depend on the mixture 
of dust grains and molecules present in a system. Tracing them through cosmic 
times improves our knowledge of the formation and evolution of dust. 
Furthermore, characterising the typical extinction curves of high-redshift 
galaxies is crucial for studies of the evolution of the SFR through cosmic 
times, in particular, if the SFR is determined from the rest-frame UV stellar 
continuum. Therefore, the number of investigations on the properties of the 
dusty ISM at intermediate/high redshifts has been increasing in recent times 
(e.g., Malhotra \cite{MAL97}; Pitman et al. \cite{PIT00}; Maiolino et al. 
\cite{MAI01}; Vernet et al. \cite{VER01}; Vijh et al. \cite{VIJ03}; Hopkins 
et al. \cite{HOP04}; Savaglio \& Fall \cite{SAVF04}; Wang et al. 
\cite{WAN04}; Wild \& Hewett \cite{WIL05}; \cite{NOL05}; York et al. 
\cite{YOR06}). In particular, \cite{NOL05} found that UV-luminous galaxies at 
$2 < z < 2.5$ can host dust-producing extinction curves with a significant UV 
bump. They made use of radiative transfer models with the same dust/stars 
configuration that helps reproducing properties of dust attenuation in 
starburst galaxies in the local Universe (Gordon et al. \cite{GOR97}) as well 
as in the high-redshift one (Vijh et al. \cite{VIJ03}). Under this 
assumption, \cite{NOL05} concluded that the objects in their sample appear to 
host a mixture of dust grains that produce extinction curves analogous to 
those of the SMC and LMC and in between them. Interestingly, Wild \& Hewett 
(\cite{WIL05}) found that damped Ly$\alpha$ absorbers with 
$0.84 < z_{\rm abs} < 1.3$ (identified through Ca\,II (H \& K) absorption) 
exhibit extinction curves that are compatible with both the LMC and SMC ones.

A possible connection between these two results is offered by our study. In 
fact, in the previous sections we have provided a robust evidence for the 
diversity of the extinction curves characterising 108 massive, UV-luminous
galaxies at $1 < z < 2.5$. Hence, the main conclusion of \cite{NOL05} is
extended down to $z \sim 1$. This evidence stems from the distribution of the
data points in the diagnostic diagram portraying the proxy for the strength of
the 2175\,\AA{} feature $\gamma_{34}$ versus the proxy for reddening in the
rest-frame UV $\beta_{\rm b}$ (Fig.~\ref{fig_gamma34_betab}). In addition to
the radiative transfer models used by \cite{NOL05}, we make use of additional
ones that describe dust attenuation for a disc geometry and include the effect
of an age-dependent extinction (Silva et al. \cite{SIL98}; Pierini et al.
\cite{PIE04}). This analysis enforces the presence of a considerable fraction
of dust out of the disc mid-plane of the sample galaxies, confirming that the
dust/stars configuration assumed by \cite{NOL05} was not too wrong,
though simple. The physical foundation of this complex dust/stars 
configuration is the existence of galactic winds (see, e.g., Murray et al. 
\cite{MUR05}). More noticeably, the evidence of dust with a UV bump also 
stems from the direct inspection of the composite spectra obtained for two 
subsamples of galaxies selected as a function of $\gamma_{34}$ either at 
$1 < z < 1.5$ (Fig.~\ref{fig_mean_sb_wb}) or at $2 < z < 2.5$ (fig.~11 in 
\cite{NOL05}).

Our analysis clearly benefits from high S/N spectroscopy, whereas others are 
limited by broad-band photometry. Furthermore, unlike other samples, ours 
does not suffer from poor statistics and/or the removal of the dominant
contribution to the emission by an AGN and/or the potential modification of 
the dusty ISM by the hard radiation field produced by AGN activity. Finally, 
it is made of galaxies detected in emission and not in absorption, which can 
thus be morphologically identified and studied at several wavelengths. These 
UV-luminous galaxies at $1 < z < 2.5$ are characterised by large total 
stellar masses and high SFRs as a consequence of the selection criteria 
(Sect.~\ref{data})\footnote{For instance, the need for spectra with a high 
S/N at rest-frame UV wavelengths implies a large fraction of young (i.e., 
$< 100$\,Myr-old) stellar populations.}. Their stellar masses and SFRs 
tend to be lower the lower the redshift, consistently with the so-called
``downsizing scenario'' (Cowie et al. \cite{COW96}; Gavazzi et al.
\cite{GAV96}; Juneau et al. \cite{JUN05}). In fact, this scenario predicts a 
decrease with cosmic age of the transition mass separating actively from 
passively-evolving galaxies (e.g., Pannella et al. \cite{PAN06}).

In our sample, galaxies that are heavily reddened in the rest-frame UV tend 
to be more massive and have higher SFRs, whatever the redshift. Furthermore, 
they preferentially exhibit an extinction curve with a significant UV bump. 
Nevertheless, we do not find any evidence that the observed strength of the 
UV bump is directly related to either stellar mass or SFR. Clearly these 
apparently conflicting results call for an interpretation that links 
properties of the stellar populations to those of the dusty ISM.

For objects at $1 < z < 1.5$ it is possible to measure the 4000\,\AA{} break
index D4000 and/or the Balmer-break proxy $C$(36-39). These indices suggest 
that dust with a significant UV bump is present in galaxies where 
intermediate-age (i.e., from $0.2$ to 1 -- 2\,Gyr-old) stellar populations 
exist and/or that are about three times older than those without a UV bump
detection (see Sect.~\ref{stelpop_gamma34}). The existence of a link between 
extinction curve and the characteristic age of the stellar populations of a 
galaxy is nicely supported by the analysis of the absorption-line ratios that 
are available for the same objects. In fact, we find that the ratio of the 
absorption strengths in the Mg\,II doublet at 2800\,\AA{} and in the Fe\,II 
doublet at 2600\,\AA{} is larger in more UV-reddened galaxies with evidence 
of a UV bump. This holds whether the absorption arises in the ISM or in 
stellar atmospheres. Finally, galaxies with extinction curves exhibiting
a significant UV bump at $1 < z < 1.5$ predominantly appear as ordinary disc 
systems, at variance with galaxies without a UV-bump detection at similar 
redshifts, which appear as irregular systems. Smoother morphologies are also 
consistent with a relatively longer star-formation history. 

In general, models with a mixed dust/stars configuration where MW-type dust 
is distributed in a geometrically thin disc (e.g., Silva et al. 
\cite{SIL98}; Pierini et al. \cite{PIE04}) can account for values of 
$-2.5 < \gamma_{34} < 1$ in our sample of $1 < z < 1.5$ galaxies (see 
Fig.~\ref{fig_dustevol_disc}). However, they predict a rapid decrease of the 
observed strength of the UV bump as soon as the attenuation at UV wavelengths 
becomes large along the line of sight. In order to describe the increase of 
the observed strength of the UV bump with increasing reddening and the 
existence of galaxies with $\gamma_{34} < -3$, a screen-like geometry needs 
to be invoked (e.g., Witt \& Gordon \cite{WIT00}). The presence of large 
amounts of dust above the disc mid-plane can be obtained in principle via 
galactic winds, starburst superwinds, and even AGN activity. It is favoured 
by the presence of magnetic fields (e.g., Greenberg et al. \cite{GRE87}). On 
the other hand, the ISM of objects with $\gamma_{34} \sim 1$ does not seem to 
contain a significant fraction of carriers of the UV bump. The dust 
attenuation properties of these galaxies are similar to those of nearby 
starbursts (e.g., Calzetti et al. \cite{CAL94}). Consistently, our 
$1 < z < 1.5$ galaxies without a detection of the UV bump tend to be 
dominated by young (i.e., $\la 100$\,Myr-old) stellar populations and to have 
irregular morphologies.

There is no spectral indication for the characteristic ages of the stellar 
populations in the UV-bright galaxies at $2 < z < 2.5$. However, the mean 
$J - K$ colour of the subsample of FDF galaxies at these redshifts 
($\langle J - K \rangle = 1.7$) suggests that the bulk of their stellar 
populations is most probably not much older than 1\,Gyr (Franx et al. 
\cite{FRA03}; van Dokkum et al. \cite{DOK04}; Pierini et al. \cite{PIE05}). 
Furthermore, their metallicities appear to be about solar, as well as for 
the galaxies at $1 < z < 1.5$ (see Sect.~\ref{stelpop_gamma34}).

On the other hand, the subsample of UV-bright galaxies at $2 < z < 2.5$
offers the possibility to probe the ISM topology from the determination of
the EWs of strong interstellar absorption lines (cf. Shapley et al.
\cite{SHA03}). For these objects, \cite{NOL05} concluded that the presence
of the UV bump does not seem to depend on the total metallicity, as given by
the EW of the C\,IV doublet at 1550\,\AA{}. Conversely, it seems to be
associated with a large average EW of the six most prominent interstellar
low-ionisation absorption lines in the far-UV. \cite{NOL05} interpreted this 
result as indication of a larger covering fraction of young massive stars by 
neutral gas clouds (containing dust grains).

At this point, it is possible to piece together the individual clues on the 
origin of the different extinction curves of UV-luminous galaxies at 
$1 < z < 2.5$ and formulate a self-consistent scenario. In particular, we 
can address physical reasons for the presence or absence of the UV bump.
First of all, we recall that the nature of the carriers of the 2175\,\AA{}
feature is probably multiple, i.e., organic carbon (as in clusters of 
polycyclic aromatic hydrocarbons, PAHs) and amorphous silicates (Bradley et 
al. \cite{BRA05} and references therein). As supernovae (SNe) manufacture 
many of the condensible elements, clearly they are candidate sources of dust. 
In particular, the expanding envelope of a type-II SN may represent an O-rich 
environment where silicate dust can form. Silicon carbide and amorphous 
carbon are the main products of C-rich stars of intermediate mass (Whittet 
\cite{WHI03}). Conversely, intermediate-mass, O-rich stars produce silicate 
dust. These stars lose mass copiously during the red giant and asymptotic 
giant branch phases of their evolution. As seen before, AGB stars (with ages 
from 0.2 to 1 -- 2\,Gyr) are most probably present in large numbers in 
galaxies which exhibit the 2175\,\AA{} absorption feature in their spectra. 
These stars are considered to be the main source of the probable carriers of 
the UV bump also in the local Universe (e.g., Galliano \cite{GAL06}).

Once the basic units are produced and injected into the general ISM, dust 
reprocessing takes place there. Our study seems to confirm that dust 
self-shielding is important for the survival of the carriers of the UV bump 
in the general ISM (cf. Gordon et al. \cite{GOR03}), despite other processes 
may modify or destroy them (see Whittet et al. \cite{WHI04}). This is 
consistent with recent studies on the formation and survival of PAH clusters 
in photodissociation regions (Rapacioli et al. \cite{RAP06} and references 
therein). Milder UV radiation fields, as found in galaxies at $1 < z < 1.5$, 
ease the need for self-shielding of course. An extreme example of this is the 
Milky Way, which has a relatively quiescent star-formation activity, a 
metal-enriched and aged ISM, and an average extinction curve with a very 
pronounced UV bump (e.g., Cardelli et al. \cite{CAR89}; see also
Fitzpatrick \& Massa \cite{FIT07} for a recent analysis of the shapes
of MW interstellar extinction curves). We recall that galaxies hosting dust 
with a significant UV bump at $1 < z < 1.5$ exhibit the heaviest UV 
reddening. In our preferred interpretation, this is at least partly due to 
the screen-like effect of dust lifted from the disc mid-plane by galactic 
winds.

\begin{figure}
\centering 
\includegraphics[width=8.8cm,clip=true]{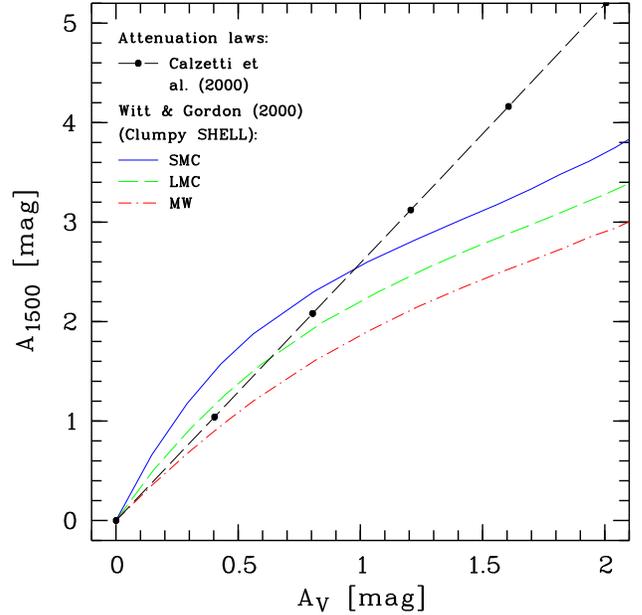}
\caption[]{Attenuation at 1500\,\AA{} versus attenuation in the $V$ band for
three radiative transfer models of Witt \& Gordon (\cite{WIT00}) with clumpy 
SHELL dust geometry and the empirical Calzetti et al. (\cite{CAL00}) law (see 
legend and Fig.~\ref{fig_betab_beta}). In the latter case filled circles are 
plotted in intervals of $\Delta E(B-V) = 0.1$.}
\label{fig_A1500_AV}
\end{figure}

Establishing the presence of different mixtures of dust grains in the dusty
interstellar media of different high-redshift galaxies has important 
consequences on the perception of the cosmological evolution of the SFR. As 
shown in Fig.~\ref{fig_A1500_AV}, attenuation models where all stars are 
enshrouded by dust predict corrections of the luminosity at 1500\,\AA{}
(rest-frame) that differ by up to $\sim 40$\% for the LMC and SMC extinction
curves at fixed attenuation in the V-band (rest-frame) $A_{\rm V} < 2$.
Furthermore, a Calzetti law systematically overpredicts the correction of 
$L_{1500}$ and, thus, of the SFR with respect to an attenuation model with 
LMC-type dust. This is also true with respect to an attenuation model with 
SMC-type dust, the overestimate being much larger than 40\% when 
$A_{\rm V} > 1$. In conclusion, not understanding amount, composition,
topology, and three-dimensional distribution of dust in a galaxy can 
seriously undermine our knowledge of galaxy evolution.

\section{Conclusions}\label{conclusions}

High signal-to-noise optical spectra available for 108 massive, UV-luminous
galaxies at $1 < z < 2.5$ allow to constrain properties of the extinction
curve at rest-frame UV wavelengths as discussed by Noll \& Pierini (2005, 
\cite{NOL05}). As a main result, we find that these galaxies host different 
mixtures of dust grains and molecules, which produce different extinction 
curves. In particular, the majority of the rest-frame UV spectral energy 
distributions (SEDs) suffering from heavy reddening exhibit a broad 
absorption excess centred at 2175\,\AA{} (the so-called ``UV bump''). Hence 
the associated galaxies host dust that can produce an extinction curve 
similar to the average one determined for the interstellar medium (ISM)
of the Large Magellanic Cloud or even of the Milky Way. Conversely, the 
extinction curve of galaxies with the least reddened UV SEDs is consistent
with the average, featureless, and steep extinction curve of the Small 
Magellanic Cloud, if the strength of the apparent UV bump is not considerably 
reduced by dust--stars mixing and age-dependent extinction effects as 
suggested for local disc galaxies.

For objects at $1 < z < 1.5$, the comparison of measured and synthetic 
stellar spectral indices shows that dust with a significant UV bump is 
present in galaxies that host a rather large fraction of intermediate-age 
stars (i.e., from $0.2$ to 1 -- 2\,Gyr old) at these redshifts. The older 
ages and the higher reddening at UV wavelengths found for these galaxies are 
consistent with the larger Mg\,II-to-Fe\,II ratio that we measure. At the 
same time, older ages are consistent with a larger fraction of smooth, 
radial surface brightness profiles in the rest-frame optical. Therefore, at 
$1 < z < 1.5$ galaxies with evidence of a UV bump appear as disc systems 
which are more evolved than the more compact, clumpy objects whose dust
properties resemble those of nearby starburst galaxies. Nevertheless, part 
of the dusty ISM seems to be lifted above the disc mid-plane owing to the 
action of galactic winds.

In a complementary way, at least for objects at $2 < z < 2.5$, the equivalent 
width of the most prominent interstellar absorption lines in the far-UV 
carries information on the topology of the ISM. As \cite{NOL05} showed, the 
covering fraction of young massive stars by neutral gas clouds (containing 
dust) is larger in galaxies with evidence of a UV bump than in those with an 
undetected 2175\,\AA{} feature. 

The previous results can be interpreted in a self-consistent way by a 
scenario where the basic units of the (probably multiple) carriers of the UV 
bump mostly originate from intermediate-mass (2 -- 5\,M$_{\odot}$) stars in 
late evolution phases. A high production rate of the carriers is, therefore,
connected to high amounts of the required stars, which needs suitable 
star-formation histories. The carriers of the 2175\,\AA{} feature are most 
probably made of organic carbon and amorphous silicates. They can survive in 
the harsh environments of UV-luminous high-redshift galaxies owing to dust
self-shielding, which is most efficient in massive, gas-rich, 
high-metallicity galaxies. As indicated by our analysis, good candidates for
galaxies with a pronounced UV bump in their extinction curves (obviously 
fulfilling the requirements above) are highly dust-reddened, UV-bright
galaxies at $2 < z < 2.5$ and near-IR bright, intensely star-forming galaxies 
at $1 < z < 1.5$.

The existence of different extinction curves implies that different patterns
of evolution and reprocessing of dust exist among massive, UV-luminous
galaxies at high redshift. Ignoring this may produce a non-negligible
uncertainty on the SFR estimate of a galaxy based on the rest-frame UV.

\begin{acknowledgements}
We thank Karl D. Gordon for providing dust attenuation models with LMC-type 
dust. We also wish to thank Anna Gallazzi and Adolf N. Witt for useful 
discussions and suggestions. MP thanks Chris Conselice for making available 
the IRAF code of the CAS algorithm. Finally, we thank the anonymous referee 
for her/his helpful comments that improved this paper. The spectra of the 
FDF and K20 samples are based on observations obtained with FORS at the ESO 
VLT, Paranal, Chile. The latter spectra as well as the Hubble ACS images of 
the CDFS were retrieved from the ESO/ST-ECF Science Archive Facility. The 
GDDS spectra stem from observations with GMOS at the Gemini North Telescope, 
Mauna Kea, USA. The investigation of our sample of local starburst galaxies 
has been based on INES data from the IUE satellite and has made use of the 
NASA/IPAC Extragalactic Database (NED) which is operated by the Jet 
Propulsion Laboratory, California Institute of Technology, under contract 
with the National Aeronautics and Space Administration. This research was 
supported by the German Science Foundation (DFG, SFB 375).
\end{acknowledgements}

\begin{table*}
\caption[]{Basic parameters, i.e., ID, world coordinates (J2000), $R$ and 
$K_{\rm s}$ magnitudes, and redshift, of the 108 sample galaxies from the 
FDF, K20, and GDDS. For the origin of the data see Sect.~\ref{data}. 
Furthermore, the main parameters of this study, $\beta_b$ and $\gamma_{34}$ 
(see Sect.~\ref{parameters}), are listed. Due to non-optimal wavelength 
ranges, $\beta_b$ could only be measured for 88 objects.}
\label{tab_basic}
\centering
\begin{tabular}{c c c c c c c c}
\hline\hline
\noalign{\smallskip}
ID & RA & DEC & $R$ & $K_{\rm s}$ & $z$ & $\beta_b$ & $\gamma_{34}$ \\
\noalign{\smallskip}
\hline
\noalign{\smallskip}
CDFS-0047  & 03 32 16.35 & -27 48 23.9 & 23.43 & 19.34 & 1.295 & 
$-0.37 \pm 0.22$ & $-4.43 \pm 0.34$ \\
CDFS-0112  & 03 32 37.63 & -27 47 44.0 & 23.00 & 19.20 & 1.097 &
$-1.20 \pm 0.10$ & $-0.83 \pm 0.11$ \\
CDFS-0154  & 03 32 21.75 & -27 47 24.6 & 23.77 & 19.68 & 1.000 & 
$-0.37 \pm 0.26$ & $-4.99 \pm 0.32$ \\
CDFS-0201  & 03 32 10.57 & -27 47 06.2 & 22.92 & 19.24 & 1.045 &
$-0.90 \pm 0.10$ & $-1.29 \pm 0.16$ \\
CDFS-0217  & 03 32 30.72 & -27 46 17.2 & 23.42 & 19.25 & 1.307 & 
$-0.19 \pm 0.21$ & $-3.69 \pm 0.27$ \\
CDFS-0226  & 03 32 11.57 & -27 46 22.9 & 22.34 & 19.35 & 1.221 & 
$-1.22 \pm 0.11$ & $-0.85 \pm 0.11$ \\
CDFS-0239  & 03 32 29.09 & -27 46 29.0 & 23.11 & 19.45 & 2.227 & 
$-2.37 \pm 0.21$ & $-0.67 \pm 0.22$ \\
CDFS-0246  & 03 32 36.45 & -27 46 31.6 & 23.04 & 19.15 & 1.047 & 
$+9.99 \pm 9.99$ & $-0.65 \pm 0.28$ \\
CDFS-0254  & 03 32 08.89 & -27 46 29.1 & 24.29 & 19.82 & 1.032 & 
$+9.99 \pm 9.99$ & $-4.75 \pm 0.41$ \\
CDFS-0271  & 03 32 34.86 & -27 46 40.4 & 23.84 & 19.37 & 1.097 & 
$-0.50 \pm 0.19$ & $-7.24 \pm 0.31$ \\
CDFS-0295  & 03 32 37.36 & -27 46 45.5 & 23.53 & 19.88 & 1.843 &
$+9.99 \pm 9.99$ & $-5.25 \pm 0.13$ \\
CDFS-0310  & 03 32 19.34 & -27 47 01.1 & 23.89 & 19.14 & 1.166 & 
$-0.48 \pm 0.11$ & $-1.21 \pm 0.16$ \\
CDFS-0331  & 03 32 24.64 & -27 44 07.8 & 22.85 & 19.68 & 1.901 & 
$-1.31 \pm 0.12$ & $+1.63 \pm 0.11$ \\
CDFS-0344  & 03 32 37.41 & -27 44 07.0 & 23.43 & 18.86 & 1.017 &
$+9.99 \pm 9.99$ & $-3.15 \pm 0.30$ \\
CDFS-0346  & 03 32 23.71 & -27 44 11.8 & 23.38 & 19.31 & 2.060 & 
$-0.82 \pm 0.09$ & $-1.10 \pm 0.13$ \\
CDFS-0384  & 03 32 35.34 & -27 44 19.3 & 23.27 & 19.87 & 1.224 & 
$-0.94 \pm 0.13$ & $-2.61 \pm 0.16$ \\
CDFS-0407  & 03 32 17.96 & -27 44 31.4 & 23.21 & 19.50 & 1.227 &
$-2.02 \pm 0.12$ & $-1.89 \pm 0.23$ \\
CDFS-0413  & 03 32 36.03 & -27 44 23.8 & 22.47 & 19.16 & 1.038 & 
$-1.41 \pm 0.05$ & $-4.24 \pm 0.07$ \\
CDFS-0416  & 03 32 35.13 & -27 44 39.6 & 23.22 & 19.58 & 1.038 & 
$+0.20 \pm 0.23$ & $-6.91 \pm 0.31$ \\
CDFS-0522  & 03 32 29.12 & -27 45 21.1 & 24.32 & 19.94 & 2.226 & 
$-1.37 \pm 0.23$ & $-0.57 \pm 0.30$ \\
CDFS-0525  & 03 32 12.47 & -27 45 30.3 & 23.32 & 19.58 & 1.375 & 
$-2.41 \pm 0.17$ & $-3.31 \pm 0.19$ \\
CDFS-0582  & 03 32 12.20 & -27 45 54.4 & 22.48 & 19.20 & 1.038 &
$-0.94 \pm 0.20$ & $-1.43 \pm 0.18$ \\
CDFS-0587  & 03 32 23.09 & -27 45 58.1 & 23.35 & 19.99 & 1.727 & 
$-1.51 \pm 0.06$ & $-1.15 \pm 0.11$ \\
CDFS-0693  & 03 32 11.00 & -27 44 59.4 & 23.47 & 19.73 & 1.553 &
$-0.65 \pm 0.10$ & $-0.13 \pm 0.24$ \\
FDF-1208   & 01 05 51.87 & -25 48 03.5 & 24.01 & 20.47 & 2.178 &
$-1.46 \pm 0.14$ & $+0.06 \pm 0.39$ \\
FDF-1473   & 01 05 52.93 & -25 45 19.2 & 23.59 & 22.74 & 1.580 &
$-1.72 \pm 0.07$ & $+1.12 \pm 0.13$ \\
FDF-1496   & 01 05 53.02 & -25 48 03.9 & 23.28 & 20.24 & 1.376 &
$-0.44 \pm 0.10$ & $-1.74 \pm 0.17$ \\
FDF-1691   & 01 05 53.71 & -25 45 28.7 & 24.16 & 21.78 & 2.344 &
$-0.85 \pm 0.16$ & $-0.60 \pm 0.23$ \\
FDF-1744   & 01 05 53.90 & -25 46 06.3 & 24.35 & 21.43 & 2.374 &
$-2.54 \pm 0.21$ & $+2.42 \pm 0.41$ \\
FDF-1823   & 01 05 54.18 & -25 43 30.5 & 22.89 & 19.80 & 1.050 &
$-1.64 \pm 0.10$ & $-2.39 \pm 0.11$ \\
FDF-1922   & 01 05 54.58 & -25 47 00.1 & 23.53 & 22.32 & 1.827 &
$-2.15 \pm 0.10$ & $+0.54 \pm 0.12$ \\
FDF-1991   & 01 05 54.83 & -25 46 16.3 & 24.64 & 21.46 & 2.334 &
$-2.46 \pm 0.41$ & $+1.29 \pm 0.79$ \\
FDF-2274   & 01 05 55.89 & -25 44 34.1 & 23.57 & 20.92 & 2.253 &
$-1.59 \pm 0.10$ & $+0.42 \pm 0.40$ \\
FDF-2295   & 01 05 55.98 & -25 43 12.3 & 23.17 & 20.01 & 1.824 &
$-1.62 \pm 0.12$ & $+0.06 \pm 0.27$ \\
FDF-2334   & 01 05 56.13 & -25 44 39.9 & 23.34 & 20.98 & 1.392 &
$-1.09 \pm 0.07$ & $-0.33 \pm 0.08$ \\
FDF-2418   & 01 05 56.44 & -25 45 12.1 & 23.38 & 19.72 & 2.332 &
$-2.06 \pm 0.20$ & $-1.94 \pm 0.22$ \\
FDF-2495   & 01 05 56.75 & -25 43 43.9 & 23.59 & 20.95 & 2.453 &
$-1.15 \pm 0.09$ & $-2.51 \pm 0.69$ \\
FDF-2636   & 01 05 57.31 & -25 44 02.2 & 23.79 & 20.05 & 2.253 &
$-0.87 \pm 0.12$ & $-4.49 \pm 0.71$ \\
FDF-3005   & 01 05 58.61 & -25 48 14.1 & 23.79 & 20.08 & 2.253 &
$-1.59 \pm 0.15$ & $-1.69 \pm 0.29$ \\
FDF-3163   & 01 05 59.17 & -25 45 38.1 & 23.55 & 20.25 & 2.441 &
$-1.00 \pm 0.16$ & $-2.27 \pm 0.82$ \\
FDF-3230   & 01 05 59.40 & -25 44 07.6 & 22.72 & 18.46 & 1.154 &
$+9.99 \pm 9.99$ & $-3.07 \pm 0.72$ \\
FDF-3300   & 01 05 59.64 & -25 46 29.9 & 24.09 & 21.53 & 2.375 &
$-2.37 \pm 0.32$ & $+1.19 \pm 0.64$ \\
FDF-3374   & 01 05 59.88 & -25 45 10.7 & 23.57 & 20.22 & 2.386 &
$-1.68 \pm 0.13$ & $-0.21 \pm 0.29$ \\
FDF-3476   & 01 06 00.24 & -25 44 55.7 & 23.20 & 20.62 & 1.438 &
$-1.53 \pm 0.05$ & $+0.08 \pm 0.07$ \\
FDF-3688   & 01 06 00.89 & -25 47 05.3 & 24.25 & 21.36 & 2.375 &
$-2.36 \pm 0.27$ & $-0.58 \pm 0.61$ \\
FDF-3781   & 01 06 01.22 & -25 46 22.2 & 23.68 & 21.30 & 1.027 &
$-1.59 \pm 0.21$ & $+0.70 \pm 0.61$ \\
FDF-3810   & 01 06 01.32 & -25 45 27.9 & 22.95 & 19.74 & 2.372 &
$-1.10 \pm 0.13$ & $-3.24 \pm 0.25$ \\
FDF-3874   & 01 06 01.51 & -25 45 46.4 & 23.54 & 20.12 & 2.483 &
$-2.27 \pm 0.18$ & $-2.73 \pm 1.00$ \\
FDF-3875   & 01 06 01.51 & -25 47 33.6 & 24.64 & 22.02 & 2.243 &
$-1.91 \pm 0.13$ & $-1.14 \pm 0.35$ \\
FDF-3958   & 01 06 01.84 & -25 44 28.8 & 23.99 & 21.69 & 2.130 &
$-2.42 \pm 0.21$ & $+1.09 \pm 0.30$ \\
FDF-4049   & 01 06 02.15 & -25 47 17.5 & 23.51 & 20.26 & 1.475 &
$-1.11 \pm 0.05$ & $-2.99 \pm 0.07$ \\
FDF-4092   & 01 06 02.33 & -25 43 29.5 & 23.77 & 22.80 & 1.394 &
$-1.58 \pm 0.09$ & $+0.00 \pm 0.11$ \\
FDF-4115   & 01 06 02.38 & -25 47 24.8 & 21.87 & 19.09 & 1.317 &
$-1.67 \pm 0.08$ & $+0.12 \pm 0.08$ \\
FDF-4310   & 01 06 03.03 & -25 43 56.9 & 23.76 & 21.30 & 1.667 &
$-1.66 \pm 0.15$ & $+0.39 \pm 0.20$ \\
FDF-4324   & 01 06 03.09 & -25 43 57.8 & 23.65 & 20.60 & 1.667 &
$-1.39 \pm 0.14$ & $+0.76 \pm 0.24$ \\
FDF-4479   & 01 06 03.70 & -25 44 53.5 & 23.56 & 20.79 & 1.304 &
$-1.02 \pm 0.07$ & $-1.09 \pm 0.12$ \\
\noalign{\smallskip}
\hline
\end{tabular}
\end{table*}

\begin{table*}
\addtocounter{table}{-1}
\caption[]{\em continued}
\label{tab_basic_sup1}
\centering
\begin{tabular}{c c c c c c c c}
\hline\hline
\noalign{\smallskip}
ID & RA & DEC & $R$ & $K_{\rm s}$ & $z$ & $\beta_b$ & $\gamma_{34}$ \\
\noalign{\smallskip}
\hline
\noalign{\smallskip}
FDF-4795   & 01 06 04.81 & -25 47 13.9 & 23.66 & 20.78 & 2.159 &
$-1.95 \pm 0.18$ & $+0.05 \pm 0.18$ \\
FDF-4871   & 01 06 05.06 & -25 46 03.5 & 23.53 & 20.53 & 2.472 &
$-1.22 \pm 0.21$ & $-3.89 \pm 0.50$ \\
FDF-4996   & 01 06 05.50 & -25 46 27.8 & 23.45 & 20.91 & 2.028 &
$-1.36 \pm 0.08$ & $-0.74 \pm 0.13$ \\
FDF-5058   & 01 06 05.72 & -25 46 26.3 & 23.48 & 21.07 & 2.027 &
$-1.79 \pm 0.07$ & $+0.84 \pm 0.11$ \\
FDF-5072   & 01 06 05.76 & -25 45 18.8 & 22.88 & 20.13 & 1.389 &
$-1.50 \pm 0.06$ & $-0.42 \pm 0.08$ \\
FDF-5135   & 01 06 05.97 & -25 44 43.8 & 23.89 & 20.76 & 2.346 &
$-1.37 \pm 0.34$ & $+0.04 \pm 0.34$ \\
FDF-5152   & 01 06 06.01 & -25 45 40.4 & 22.89 & 20.22 & 1.370 &
$-1.32 \pm 0.05$ & $-0.28 \pm 0.06$ \\
FDF-5165   & 01 06 06.06 & -25 44 43.4 & 23.53 & 20.39 & 2.346 &
$-2.20 \pm 0.26$ & $-0.14 \pm 0.37$ \\
FDF-5182   & 01 06 06.13 & -25 45 33.7 & 23.83 & 21.52 & 1.369 &
$-1.78 \pm 0.09$ & $-0.39 \pm 0.10$ \\
FDF-5190   & 01 06 06.15 & -25 44 43.1 & 24.66 & 22.78 & 2.347 &
$-1.36 \pm 0.41$ & $-1.53 \pm 0.38$ \\
FDF-5227   & 01 06 06.29 & -25 43 51.8 & 24.09 & 21.75 & 2.399 &
$-1.34 \pm 0.32$ & $-4.83 \pm 0.36$ \\
FDF-5236   & 01 06 06.34 & -25 45 42.2 & 22.54 & 18.94 & 1.153 &
$-1.69 \pm 0.15$ & $-1.35 \pm 0.18$ \\
FDF-5458   & 01 06 07.18 & -25 46 44.2 & 23.52 & 20.35 & 1.513 &
$-1.30 \pm 0.07$ & $-1.00 \pm 0.13$ \\
FDF-5585   & 01 06 07.65 & -25 45 50.6 & 21.78 & 17.79 & 1.074 &
$+9.99 \pm 9.99$ & $-1.37 \pm 0.52$ \\
FDF-5667   & 01 06 07.94 & -25 46 37.3 & 23.69 & 20.61 & 1.024 &
$+9.99 \pm 9.99$ & $-1.36 \pm 1.59$ \\
FDF-6024   & 01 06 09.23 & -25 48 14.1 & 22.27 & 19.85 & 2.372 &
$-2.04 \pm 0.17$ & $-1.28 \pm 0.36$ \\
FDF-6101   & 01 06 09.50 & -25 45 50.0 & 22.96 & 20.14 & 1.142 &
$-0.88 \pm 0.13$ & $-2.34 \pm 0.18$ \\
FDF-6344   & 01 06 10.36 & -25 43 33.4 & 23.32 & 22.18 & 1.663 &
$-1.88 \pm 0.06$ & $-0.81 \pm 0.10$ \\
FDF-6358   & 01 06 10.39 & -25 44 30.8 & 23.19 & 19.77 & 1.315 &
$-0.64 \pm 0.18$ & $-0.32 \pm 0.21$ \\
FDF-6372   & 01 06 10.45 & -25 48 29.0 & 23.55 & 20.86 & 2.349 &
$-1.73 \pm 0.22$ & $+0.14 \pm 0.19$ \\
FDF-6384   & 01 06 10.49 & -25 45 13.0 & 23.18 & 20.69 & 1.314 &
$-1.08 \pm 0.07$ & $+0.71 \pm 0.08$ \\
FDF-6407   & 01 06 10.57 & -25 45 31.7 & 23.95 & 20.70 & 2.162 &
$-1.36 \pm 0.12$ & $+1.62 \pm 0.32$ \\
FDF-6432   & 01 06 10.65 & -25 46 09.2 & 23.70 & 19.71 & 1.153 &
$-0.11 \pm 0.27$ & $-1.04 \pm 0.34$ \\
FDF-6547   & 01 06 11.06 & -25 46 49.1 & 23.74 & 21.27 & 1.212 &
$-1.87 \pm 0.08$ & $-0.42 \pm 0.09$ \\
FDF-6864   & 01 06 12.19 & -25 43 56.2 & 23.61 & 20.93 & 1.394 &
$-1.57 \pm 0.06$ & $-0.34 \pm 0.08$ \\
FDF-6934   & 01 06 12.36 & -25 44 56.8 & 23.08 & 20.16 & 2.445 &
$-2.05 \pm 0.26$ & $-0.09 \pm 0.92$ \\
FDF-6947   & 01 06 12.40 & -25 48 14.8 & 24.13 & 20.57 & 2.357 &
$-1.87 \pm 0.16$ & $-3.46 \pm 0.19$ \\
FDF-7029   & 01 06 12.67 & -25 45 58.4 & 23.86 & 21.52 & 2.374 &
$-1.91 \pm 0.21$ & $-2.26 \pm 0.27$ \\
FDF-7078   & 01 06 12.80 & -25 46 00.5 & 24.15 & 21.97 & 2.378 &
$-1.47 \pm 0.35$ & $+2.70 \pm 0.98$ \\
FDF-7223   & 01 06 13.30 & -25 44 32.7 & 23.27 & 20.12 & 1.051 &
$+9.99 \pm 9.99$ & $-3.69 \pm 2.20$ \\
FDF-7307   & 01 06 13.60 & -25 47 25.1 & 24.28 & 21.06 & 2.438 &
$-2.00 \pm 0.33$ & $+2.19 \pm 0.86$ \\
FDF-7342   & 01 06 13.70 & -25 46 13.2 & 24.07 & 20.70 & 2.375 &
$-1.55 \pm 0.17$ & $-4.14 \pm 0.79$ \\
FDF-7345   & 01 06 13.71 & -25 46 23.6 & 23.62 & 20.78 & 1.025 &
$-1.95 \pm 0.08$ & $-1.11 \pm 0.11$ \\
FDF-8604   & 01 06 18.61 & -25 48 17.5 & 21.53 & 18.04 & 1.142 &
$-0.03 \pm 0.1$7 & $-5.74 \pm 0.22$ \\
Q0055-0017 & 00 57 58.22 & -26 40 59.2 & 23.84 & 19.68 & 1.128 & 
$-0.91 \pm 0.13$ & $-4.17 \pm 0.31$ \\
Q0055-0023 & 00 57 56.80 & -26 41 04.7 & 23.06 & 19.87 & 1.436 &
$-1.04 \pm 0.12$ & $-2.07 \pm 0.15$ \\
Q0055-0024 & 00 57 54.72 & -26 41 07.5 & 22.33 & 19.94 & 1.374 & 
$-1.19 \pm 0.04$ & $-0.06 \pm 0.08$ \\
Q0055-0050 & 00 58 05.67 & -26 41 27.0 & 23.66 & 19.25 & 1.002 & 
$+9.99 \pm 9.99$ & $-5.74 \pm 0.59$ \\
Q0055-0052 & 00 58 07.03 & -26 41 27.0 & 21.88 & 18.99 & 2.175 & 
$-0.67 \pm 0.31$ & $-0.49 \pm 0.34$ \\
Q0055-0119 & 00 58 00.34 & -26 42 26.5 & 23.20 & 19.37 & 1.059 & 
$-0.11 \pm 0.13$ & $-3.27 \pm 0.20$ \\
Q0055-0131 & 00 57 59.39 & -26 42 37.2 & 22.93 & 19.42 & 1.367 & 
$-0.94 \pm 0.08$ & $-0.52 \pm 0.12$ \\
Q0055-0157 & 00 58 07.79 & -26 42 52.9 & 22.33 & 19.91 & 1.314 & 
$+9.99 \pm 9.99$ & $+1.21 \pm 0.05$ \\
Q0055-0255 & 00 57 50.82 & -26 43 52.5 & 23.47 & 18.75 & 1.105 & 
$+9.99 \pm 9.99$ & $-0.62 \pm 0.27$ \\
Q0055-0262 & 00 58 08.42 & -26 43 55.4 & 23.30 & 19.45 & 1.075 & 
$+9.99 \pm 9.99$ & $+0.15 \pm 0.22$ \\ 
SA02-1417  & 02 09 33.32 & -04 37 31.2 & 23.92 & 19.39 & 1.599 &
$+9.99 \pm 9.99$ & $-0.69 \pm 0.32$ \\
SA12-6301  & 12 05 18.36 & -07 23 43.3 & 22.70 & 19.49 & 1.760 &
$+9.99 \pm 9.99$ & $-1.17\pm  0.10$ \\
SA12-6339  & 12 05 32.70 & -07 23 37.7 & 23.71 & 20.15 & 2.293 &
$+9.99 \pm 9.99$ & $+0.96 \pm 0.12$ \\
SA12-7250  & 12 05 32.88 & -07 22 32.8 & 23.98 & 19.94 & 1.900 &
$+9.99 \pm 9.99$ & $+0.19 \pm 0.24$ \\
SA15-4762  & 15 23 53.27 & -00 06 42.0 & 24.39 & 19.86 & 1.598 &
$+9.99 \pm 9.99$ & $-7.11 \pm 0.37$ \\
SA15-6396  & 15 23 41.04 & -00 05 10.8 & 23.92 & 21.00 & 1.928 &
$+9.99 \pm 9.99$ & $+1.12 \pm 0.09$ \\
SA15-6488  & 15 23 45.55 & -00 05 05.2 & 24.13 & 21.00 & 2.044 &
$+9.99 \pm 9.99$ & $+1.01 \pm 0.22$ \\
SA15-7353  & 15 23 40.43 & -00 03 54.1 & 24.68 & 19.89 & 2.091 &
$+9.99 \pm 9.99$ & $+0.06 \pm 0.25$ \\
\noalign{\smallskip}
\hline
\end{tabular}
\end{table*}

\begin{table*}
\caption[]{Average properties of the six subsamples defined in 
Sect.~\ref{globpar_gamma34}. The numbers of objects $N$ used for the 
derivation of the morphological parameters (in parentheses) are smaller than 
for the complete subsamples, since Hubble ACS images are not available for 
the K20 quasar field (Sect.~\ref{K20}) and some outer parts of the FDF 
(Sect.~\ref{FDF}). In the following we list the magnitudes $R$ and 
$K_{\rm s}$, the UV-continuum slope parameter $\beta_{\rm b}$, the UV-bump 
tracer $\gamma_{34}$, the luminosity at 1500\,\AA{} $L_{1500}$, the dust 
reddening parameter $E(B-V)$, the bolometric luminosity $L_{\rm bol}$ 
and the SFR from the model fits, the total stellar mass $M_{\rm stars}$, the 
specific SFR $\phi$, the effective radius $R_{\rm e}$, the S\'ersic index 
$n_{\rm ser}$, the ellipticity $e$, the CAS parameters concentration $C$ and 
asymmetry $A$, and the ``clumpiness'' $R_{\rm T} + R_{\rm A}$.}
\label{tab_avpar}
\centering
\begin{tabular}{c c c c c c c}
\hline\hline
\noalign{\smallskip}
Parameter                          & 
$1 < z < 1.5$          & $1 < z < 1.5$          & 
$1 < z < 1.5$          & $2 < z < 2.5$  & $2 < z < 2.5$      & 
$2 < z < 2.5$      \\
                                   & 
$\beta_{\rm b} < -1.5$ & $\beta_{\rm b} > -1.5$ & 
$\beta_{\rm b} > -1.5$ & $\beta < -0.4$ & $\beta > -0.4$     & 
$\beta > -0.4$     \\
                                   &
                       & $\gamma_{34} > -2$     &  
$\gamma_{34} < -2$     &                & $\gamma_{34} > -2$ & 
$\gamma_{34} < -2$ \\ 
\noalign{\smallskip}
\hline
\noalign{\smallskip}
$N$                                & 
13(12)           & 14(12)           & 13(9)            & 
16               & 9                & 9                \\
$R$ [mag]                          &
$23.44 \pm 0.11$ & $23.17 \pm 0.09$ & $23.21 \pm 0.17$ &
$23.89 \pm 0.15$ & $23.77 \pm 0.13$ & $23.69 \pm 0.13$ \\
$K_{\rm s}$ [mag]                  &
$20.94 \pm 0.24$ & $20.24 \pm 0.11$ & $19.73 \pm 0.09$ &
$21.22 \pm 0.18$ & $20.66 \pm 0.22$ & $20.52 \pm 0.20$ \\
$\beta_{\rm b}$                    &
$-1.70 \pm 0.06$ & $-0.95 \pm 0.11$ & $-0.72 \pm 0.10$ &
$-1.96 \pm 0.09$ & $-1.61 \pm 0.16$ & $-1.37 \pm 0.15$ \\ 
$\gamma_{34}$                      &
$-0.44 \pm 0.18$ & $-0.63 \pm 0.29$ & $-3.70 \pm 0.31$ &
$+0.36 \pm 0.36$ & $-0.44 \pm 0.35$ & $-3.50 \pm 0.30$ \\
$\log L_{1500}$ [W/\AA{}]          &
$33.69 \pm 0.07$ & $33.67 \pm 0.06$ & $33.48 \pm 0.07$ &
$34.10 \pm 0.05$ & $34.06 \pm 0.06$ & $34.17 \pm 0.05$ \\
$E_{\rm B-V}$                      &
 $0.26 \pm 0.01$ &  $0.38 \pm 0.02$ &  $0.42 \pm 0.02$ & 
 $0.27 \pm 0.01$ &  $0.37 \pm 0.02$ &  $0.39 \pm 0.02$ \\
$\log L_{\rm bol}$ [W/\AA{}]       &
$11.55 \pm 0.09$ & $12.02 \pm 0.06$ & $11.99 \pm 0.07$ &
$11.97 \pm 0.08$ & $12.39 \pm 0.08$ & $12.56 \pm 0.08$ \\
$\log {\rm SFR}$ [M$_{\odot}$/yr]  &
 $1.83 \pm 0.08$ &  $2.30 \pm 0.08$ &  $2.23 \pm 0.09$ &
 $2.24 \pm 0.09$ &  $2.61 \pm 0.08$ &  $2.82 \pm 0.08$ \\
$\log M_{\rm stars}$ [M$_{\odot}$] &
 $9.91 \pm 0.21$ & $10.18 \pm 0.09$ & $10.40 \pm 0.06$ &
$10.14 \pm 0.11$ & $10.67 \pm 0.11$ & $10.76 \pm 0.13$ \\
$\log \phi$ [Gyr$^{-1}$]           &
 $0.89 \pm 0.17$ &  $1.11 \pm 0.14$ &  $0.79 \pm 0.16$ &
 $1.06 \pm 0.13$ &  $0.93 \pm 0.11$ &  $1.06 \pm 0.14$ \\
$R_{\rm e}$ [kpc] (from $\log R_{\rm e}$) &
 $2.85 \pm 0.43$ &  $3.59 \pm 0.33$ &  $3.99 \pm 0.52$ &
 $2.79 \pm 0.52$ &  $3.73 \pm 1.36$ &  $3.83 \pm 0.82$ \\
$n_{\rm ser}$                      &
 $1.32 \pm 0.36$ &  $0.98 \pm 0.23$ &  $0.54 \pm 0.25$ &
 $2.07 \pm 0.40$ &  $2.48 \pm 0.47$ &  $1.61 \pm 0.49$ \\
$e$                                &
 $0.49 \pm 0.07$ &  $0.42 \pm 0.06$ &  $0.45 \pm 0.07$ &
 $0.63 \pm 0.04$ &  $0.49 \pm 0.07$ &  $0.62 \pm 0.07$ \\
$C_{\rm CAS}$                      &
 $2.43 \pm 0.09$ &  $2.16 \pm 0.11$ &  $2.07 \pm 0.11$ &
 $2.52 \pm 0.11$ &  $2.46 \pm 0.12$ &  $2.19 \pm 0.17$ \\
$A_{\rm CAS}$                      &
 $0.20 \pm 0.06$ &  $0.19 \pm 0.04$ &  $0.28 \pm 0.08$ &
 $0.18 \pm 0.02$ &  $0.18 \pm 0.05$ &  $0.22 \pm 0.05$ \\
$R_{\rm T} + R_{\rm A}$            &
 $0.24 \pm 0.02$ &  $0.24 \pm 0.03$ &  $0.26 \pm 0.04$ &
 $0.25 \pm 0.05$ &  $0.40 \pm 0.09$ &  $0.32 \pm 0.05$ \\
\noalign{\smallskip}
\hline
\end{tabular}
\end{table*}

\end{document}